\RequirePackage{fix-cm}
\documentclass[smallextended]{svjour3}       

\usepackage{enumitem}
\usepackage{hyperref}
\usepackage{indentfirst}
\usepackage{multirow}%
\usepackage{amsmath,amssymb,amsfonts}%
\usepackage{mathrsfs}%
\usepackage[title]{appendix}%
\usepackage{textcomp}%
\usepackage{manyfoot}%
\usepackage{booktabs}%
\usepackage{subfig}%
\usepackage{array}
\usepackage{algorithm}%
\usepackage{algorithmicx}%
\usepackage{algpseudocode}%
\usepackage{listings}%
\usepackage{tikz}
\usepackage{csquotes}
\usepackage{threeparttable}
\usepackage{graphicx} 
\usepackage[many]{tcolorbox}    	
\usepackage{makecell}
\usepackage{threeparttable}
\usepackage{adjustbox}
\usepackage{siunitx}
\usepackage{bookmark}
\usepackage{xcolor}
\usepackage{xcolor}
\usepackage{marginnote}

\begin{document}


\title{The Dangers of Non--Self-Fixed Architecture Technical Debt and Its Impact on Time-to-Fix}


\author{Edi Sutoyo \and
        Paris Avgeriou \and 
        Andrea Capiluppi
} 
\authorrunning{E. Sutoyo et al.}
%
\institute{E. Sutoyo \at 
            1. Bernoulli Institute, University of Groningen, Groningen, The Netherlands\\
            2. Department of Information Systems, Telkom University, Bandung, Indonesia\\
            \email{e.sutoyo@rug.nl}
            \and
            P. Avgeriou \at
            Bernoulli Institute, University of Groningen, Groningen, The Netherlands\\
            \email{p.avgeriou@rug.nl}
            \and
            A. Capiluppi \at
            Bernoulli Institute, University of Groningen, Groningen, The Netherlands\\
            \email{a.capiluppi@rug.nl}
            }

\date{Received: date / Accepted: date}

\maketitle              

\abstract{
Technical Debt (TD) refers to the long-term costs incurred when developers prioritize short-term delivery over quality-improving work. Architectural Technical Debt (ATD) arises when architectural decisions (e.g., technology choices, patterns, or decomposition) prioritize near-term progress over future maintainability and evolvability. Because ATD affects a system's core structure and propagates through architectural dependencies, it is often more expensive and disruptive to remediate than localized code-level debt. Although ATD has been widely studied, an important but underexplored aspect of repayment is who performs it. Prior work provides limited empirical evidence on repayment responsibility in ATD and its relationship to time-to-fix.

We empirically study self-fixed ATD, where the introducer also repays the debt, and contrast it with non--self-fixed ATD in large Apache open-source projects. We reconstruct ATD lifecycles by tracing Jira artifacts to version-control history to identify introduction and repayment points and attribute developer roles. We address three research questions on the prevalence of self-fixed ATD, time-to-fix differences between self-fixed and non--self-fixed items, and how factors related to code change and collaboration metrics relate to repayment speed. Using descriptive statistics, non-parametric tests, and survival analysis, we show that self-fixed and non--self-fixed ATD exhibit distinct repayment dynamics and differences in how changes are shared on ATD-affected files. In particular, non--self-fixed ATD is more likely to remain unresolved longer when changes are spread across many developers. These results provide actionable guidance for maintainers to identify potentially high-risk ATD items and support knowledge transfer by maintaining introducer involvement when feasible and documenting the design rationale during repayment.
}
\keywords{empirical study, self-fixed, architecture technical debt, technical debt}
\maketitle


\section{Introduction}\label{secIntro}
Technical Debt (TD) is a well-established metaphor describing the long-term costs that arise when developers adopt expedient solutions or defer quality-improving work to achieve short-term delivery goals \cite{cunningham1992wycash,sutoyo2026self}. Like financial debt, it accrues \enquote{interest} over time (additional maintenance effort) and eventually requires repayment to restore and sustain software quality. As TD accumulates, developers typically face higher maintenance effort, slower feature delivery, and increased risk of defects and rework \cite{kruchten2012technical,li2015systematic}.

Architecture Technical Debt (ATD) is a form of TD that arises when architectural decisions, such as technology choices, patterns, or decomposition strategies, favor short-term progress at the expense of long-term maintainability \cite{martini2014architecture,besker2017impact}. Because ATD affects the system's core structure and propagates through architectural dependencies, it is more expensive and disruptive to remediate than localized code-level debt, requiring substantial architectural knowledge, non-trivial refactoring, and cross-team coordination \cite{besker2018managing,verdecchia2020architectural,martini2015investigating}. Prior work shows that ATD can hinder developers’ daily work and that its risks and costs are frequently underestimated \cite{besker2016systematic,lenarduzzi2021systematic}, making its repayment an important concern for long-term software evolution \cite{besker2017impact,verdecchia2018architectural}.

Technical debt management (TDM) encompasses activities for managing existing or potential debt throughout its lifecycle. Li et al.~\cite{li2015systematic} classify TDM into eight activities: identification, measurement, prioritization, prevention, monitoring, repayment, representation/documentation, and communication. Among these activities, our study focuses specifically on \textit{repayment}, particularly on who performs it, how long repayment takes, and the development conditions associated with different repayment modes.

Within ATD repayment, an underexplored dimension is who performs the repayment. Prior work introduced self-fixed TD, where the developer who introduced the debt also repays it, in contrast to non–self-fixed TD, where repayment is performed by another developer \cite{tan2020empirical}. Accordingly, we distinguish whether repaid ATD is handled by the same developer who introduced it or by another developer, and examine whether these repayment modes are associated with differences in repayment dynamics, particularly time-to-fix and developer involvement during the ATD lifetime. Introducers may retain contextual knowledge about the underlying decisions and constraints, which can influence repayment efficiency. This distinction matters because architectural decisions often depend on context, constraints, and design rationale that may not be fully represented in code or issue reports. When repayment is handed to another developer, some of this knowledge may need to be reconstructed or transferred, potentially increasing coordination and delaying repayment. For project teams, understanding this relationship can help identify when ATD repayment may require additional knowledge-transfer or coordination support.

However, being the introducer does not imply being the best or designated developer to repay the debt. ATD may be introduced intentionally under delivery, technical, or budget constraints; responsibilities may also change across developers, architects, and teams. Some debt may remain unrepaid because repayment is not prioritized or funded. Accordingly, we treat self-fixing as a descriptive characteristic of repayment rather than a prescription that introducers should fix their own debt. Our empirical scope is limited to repaid ATD items and examines whether repayment by the introducer or by another developer is associated with different repayment dynamics.

Tan et al.~\cite{tan2022does} studied self-fixing across code, defect, design, documentation, and test debt and found substantial self-fixing, with higher rates for code and defect debt. They also reported no significant difference in time-to-fix between self-fixed and non–self-fixed issues. Whether these findings generalize to ATD remains unclear because architectural debt often involves system-level reasoning, dependencies, and coordinated changes across multiple components \cite{martini2015investigating,samarthyam2016refactoring}. To the best of our knowledge, this is the first study to explicitly examine self-fixed ATD and compare repayment by introducers and other developers in terms of prevalence and time-to-fix.

To address this gap, we trace ATD items from introduction to repayment in open-source projects and contrast self-fixed and non--self-fixed cases. Specifically, we (i) quantify how often ATD is repaid by its introducer, (ii) compare time-to-fix between self-fixed and non--self-fixed ATD, and (iii) examine how development factors, including code-change characteristics and developer involvement, differ between these repayment modes and relate to repayment time. In doing so, we characterize both who repays ATD and the development conditions associated with its repayment.

The key contributions of this study are as follows:
\begin{itemize}
\item \textbf{Empirical characterization of repayment responsibility in ATD.}
We quantify how often ATD is repaid by its introducers versus by other developers across Apache open-source projects. 
\item \textbf{Comparative time-to-fix analysis of self-fixed and non--self-fixed ATD.}
We compare repayment latency using survival analysis and non-parametric statistical tests.
\item \textbf{Evidence on development factors associated with repayment mode and repayment speed.}
We examine how code-change characteristics and development activity differ between self-fixed and non--self-fixed ATD and how they relate to time-to-fix.
\end{itemize}

The rest of the study is organized as follows. Section~\ref{secRelatedWork} summarizes the related work. In Section~\ref{secStudyDesign}, we outline the study design. Section~\ref{secResults} reports the results of our empirical evaluation, and Section~\ref{secDiscussion} discusses the findings and their implications. Finally, Section~\ref{secConclusion} concludes our paper.

\section{Related Work}\label{secRelatedWork}
Prior work has examined how developers' contribution behavior and responsibilities influence software quality and maintenance outcomes. In particular, research distinguishes between maintenance performed by the \enquote*{original contributor} and maintenance performed by \enquote*{other developers} and studies how these modes differ in timing, coordination demands, and change characteristics. While the broader TD literature has investigated repayment and remediation, to the best of our knowledge, this study is the first to provide a focused empirical analysis of who repays ATD and to explicitly quantify and characterize self-fixed ATD in real-world projects. 

To position our study, this section reviews prior literature from two adjacent streams: first, we summarize findings from corrective maintenance on the relationship between developer involvement over time and latency. Second, we synthesize prior work on TD repayment and self-fixing, highlighting what is known about repayment responsibility and what remains unclear for ATD.

\paragraph{\textbf{Corrective Maintenance and Repayment\\}}
Prior studies on bug and defect repair provide the closest empirical analogy to self-fixed ATD. In this line of work, researchers distinguish between fixes performed by the original contributor (\enquote{self-fixes}) and fixes performed by other developers, showing that these modes can differ in both prevalence and repair dynamics. For example, Zhu and Godfrey~\cite{zhu2021mea} studied single-statement bug fixes and found that nearly half of simple bugs are fixed by a \emph{different} developer (44.3\%). They further observed a pronounced temporal contrast, in which fixes by the original author tend to occur much faster than those by other developers.

Tracing-based evidence reinforces this perspective. Using an SZZ-style setup on Mozilla's \texttt{comm-central}, Izquierdo-Cort\'azar et al.~\cite{izquierdo2011developers} showed that the bug-fixing committer is typically different from the bug-seeding committer(s). In particular, among fixes whose affected lines were previously modified by at most one committer (about 60\% of the sample), only around 6\% are performed by that same committer; the remainder are fixed by someone else. Even when extending the analysis to cases with up to two prior committers, most fixes are still carried out by a different developer, further indicating that self-fixing is not the dominant pattern in those settings.

A closely related stream examines how ownership, authorship, and developer experience relate to defects and quality outcomes. Rahman and Devanbu~\cite{rahman2011ownership} conducted a fine-grained study of authorship and showed that ownership/experience signals are meaningfully associated with defects. The authors argued that there is a need for analyzes that go beyond coarse, file-level ownership and focus rather on who has actually shaped the affected code. Complementing this view, Bird et al.~\cite{bird2011don} showed that ownership structures influence quality outcomes and developer behavior around code changes: that result reinforced the premise that responsibility and familiarity with code are important explanatory factors when studying who performs fixes.

Beyond defects tracked in issue systems, behavioral studies of fault insertion and fault fixing further emphasize that maintenance work is frequently performed by developers other than those who introduced the faults. In Apache Software Foundation projects, Ortu et al.~\cite{ortu2023fault} reported that fault-fixing often involves developers different from those who inserted faults and highlighted how developer experience and participation patterns shape insertion and repair behaviors over time. 

These findings also suggest that being the original contributor and having strong project-specific experience are distinct characteristics. An experienced developer may introduce an architectural change shortly after joining a project, while long-standing contributors may have better familiarity with the system. Thus, although an introducer may retain knowledge of the rationale behind a change, another developer may possess stronger project-specific knowledge. Therefore, it cannot be assumed a priori that the introducer is also the most appropriate developer to repay ATD.

The question of \enquote{who fixes} is closely related to \enquote{how long do fixes take} and where time is spent in the process. Zhang et al.~\cite{zhang2012empirical} explicitly analyzed delays around code changes by using Mylyn logs, showing that delays before and after the change are major contributors to the overall bug-fixing process time (e.g., the median delay before a change is 210.79 hours, while the median delay after a change is 8.55 hours). They also identified important factors across bug report attributes, source code properties, and code change characteristics, providing the motivation for models that treat repair time as a socio-technical rather than purely code-centric construct. 

Evidence from security maintenance further supports the presence of self-fixing in practice. Forootani et al.~\cite{forootani2022exploratory} investigated self-fixed software vulnerabilities in open-source \texttt{C} and \texttt{PHP} projects and showed that only a small portion of vulnerability fixes are performed by the same developer who introduced the vulnerability. They also found that self-fixed vulnerabilities tend to be repaired faster than non--self-fixed ones in several vulnerability categories, although the effect varies by weakness type. These results suggest that self-fixing behavior and its time-to-fix benefits are highly context-dependent, further motivating the need to examine whether similar patterns hold for architectural debt, where repayment often requires cross-component refactoring and coordination.

Many software engineering (SE) studies have prioritized curated ground truth and traceability across artifacts. For example, Wen et al. \cite{wen2019exploring} analyzed 333 bugs across seven OSS projects by linking bug-inducing and bug-fixing commits, and Al-Fraihat et al. \cite{al2024detecting} used 573 commits to classify refactoring types based on commit message text. Similarly, Prenner and Robbes \cite{prenner2021making} discussed the prevalence of small datasets in SE and contributed a benchmark dataset of 636 Jira issues, while Tsantalis et al. \cite{tsantalis2018accurate} constructed a refactoring oracle from 538 commits to enable rigorous validation of detection approaches.

These examples also highlight that small, curated datasets are common in empirical SE when traceability and ground-truth validation are central to the study design. In our context, reconstructing ATD lifecycles and attributing the \enquote*{introducer} and \enquote*{repayer} roles require linking issue-tracker and version-control artifacts with careful validation, which similarly constrains a dataset's size (see Section \ref{subsecRM} for dataset details).

\paragraph{\textbf{TD and Repayment\\}}
Self-Admitted Technical Debt (SATD) represents a form of technical debt that developers explicitly acknowledge in software-development artifacts \cite{potdar2014exploratory}. Prior studies have examined the introduction, evolution, and removal of SATD over software history, providing an important empirical basis for understanding how technical debt evolves over time~\cite{bhatia2026empirical,muse2022fixme}.

Self-fixing has been studied in the broader technical debt literature. Tan et al.~\cite{tan2020empirical} conducted the first empirical study focusing on self-fixed TD, which refers to cases where the developer who introduced the debt also fixed it later. They analyzed more than 17,000 commits from 20 Python projects in the Apache Software Foundation, using SonarQube to detect five TD types: code, defect, design, documentation, and test debt. In a follow-up study including Java projects, they found that self-fixing is more common in Python than in Java: about two-thirds of fixed issues are self-fixed in Python, compared to about one-third in Java~\cite{tan2022does}. However, SonarQube-based detection can suffer from false positives~\cite{lenarduzzi2020sonarqube}, which may inflate the observed TD prevalence and potentially bias self-fixing rates and time-to-fix estimates derived from detected issues.

To trace the TD lifecycle, Tan et al.~\cite{tan2022does} linked detected issues to commits and estimated their introduction moments using \texttt{git log} and \texttt{git blame}, complemented by manual verification. This tracing is accurate but time-consuming and may be affected by file renames and code movement. Their results show that self-fixing is frequent for TD detected by static-analysis rules, while no significant overall difference in survival time was found between self-fixed and non--self-fixed TD.

Beyond repayment prevalence and survival time, Tan et al.~\cite{tan2022does} also examined development and developer-related characteristics, including developer participation, modification frequency, seniority, file count, commit count, and commit size. Together with the maintenance studies discussed above, this work motivates our consideration of change characteristics, development activity, and project-specific developer experience. We therefore define these dimensions a priori for RQ3, and provide their precise operationalization in Section~\ref{subsubsectionRQ3}.

However, these findings do not directly address architecture technical debt. ATD typically reflects system-level design trade-offs that span modules and dependency boundaries, requiring broader architectural knowledge and coordination to repay than many code-level TD issues. Therefore, it remains unclear whether the prevalence and repayment dynamics of self-fixing observed for code-level TD generalize to ATD.

Taken together, prior work shows that who performs maintenance is related to repair dynamics across bugs, faults, vulnerabilities, and code-level TD. However, important questions remain for ATD. In particular, it is unclear how coordination during the debt lifetime relates to self-fixing and repayment time: whether repayment remains concentrated around the introducer or fixer, whether work becomes distributed across other developers, and whether such dispersion is associated with longer repayment. These questions are especially relevant for ATD because repayment may span multiple components and require architectural knowledge to be transferred or reconstructed across developers. Our study addresses this gap by examining not only who repays ATD, but also how developer involvement and development activity during the introduction--repayment interval relate to repayment mode and time-to-fix.



\section{Study Design}\label{secStudyDesign}
\subsection{Objective and Research Questions}\label{subsecRQs}

To address the identified gap, we conduct an empirical study of self-fixed ATD and its lifecycle in open-source projects. Within the broader TDM process introduced in Section~\ref{secIntro}, this study focuses on the repayment activity and, specifically, on repayment responsibility and repayment dynamics. Our objective is threefold: (i) to quantify the prevalence of self-fixed ATD by identifying how often ATD is repaid by its introducers versus by other developers, (ii) to compare repayment latency by analyzing time-to-fix between self-fixed and non--self-fixed ATD, and (iii) to examine how predefined development factors during the ATD lifetime, including code-change characteristics, development activity, and project-specific developer experience, differ across repayment modes and relate to repayment time. Based on these objectives, we formulate the following research questions (RQs):

\textbf{RQ1:} \textit{To what extent do developers self-fix ATD issues in software projects?}\\
\textbf{\textit{Rationale:}} Prior maintenance research suggests that original contributors may retain contextual knowledge about the changes they introduced, while repayment by other developers may require additional knowledge transfer and coordination. This RQ therefore examines how often the introducer performs ATD repayment, rather than assuming the introducer is necessarily the most appropriate fixer.

\textbf{RQ1.1:} \textit{How do commit- and change-size-based contribution shares on ATD-affected files differ between self-fixed and non--self-fixed ATD?}\\
\textbf{\textit{Rationale:}} RQ1 measures \emph{whether} ATD is self-fixed, but it does not show \emph{how} work is distributed on the affected files during the item's lifetime. We therefore examine per-file contribution shares using two complementary measures: commit frequency and change size (added plus deleted lines).

\textbf{RQ2:} \textit{What is the time span between the introduction and repayment of self-fixed ATD compared to non--self-fixed ATD?}\\
\textbf{\textit{Rationale:}} The key concern behind non--self-fixing ATD is whether repayment performed by developers other than the introducer affects time-to-fix. 

\textbf{RQ2.1:} \textit{For non--self-fixed ATD items, how does time-to-fix vary with developers' involvement during the interval from introduction to repayment?}\\
\textbf{\textit{Rationale:}} For non--self-fixed ATD, repayment is done by someone other than the introducer, and repayment speed may depend on how involved the introducer remains before the fix. We therefore examine whether time-to-fix varies with introducer and fixer involvement on the affected files during the introduction--repayment interval.


\textbf{RQ3:} \textit{How do development factors differ between self-fixed and non--self-fixed ATD and relate to time-to-fix?}\\
\textbf{\textit{Rationale:}} If the introducers are not involved in fixing ATD, recovering the architectural rationale and coordinating changes may require more effort, which may be reflected in change scope, the number of commits, the number of involved developers, and how contributions are distributed across developers. We therefore compare these factors between self-fixed and non--self-fixed ATD and examine how they relate to time-to-fix.

\subsection{Research Methodology}\label{subsecRM}
To investigate the ATD lifecycle, we conduct a repository-based empirical study integrating Jira issue-tracker and version-control data. The study design follows the empirical case-study guidelines of Runeson et al.~\cite{runeson2012case}, consistent with the methodological approach adopted by Tan et al.~\cite{tan2022does}. We next describe the lifecycle moments considered in the study, the data used to reconstruct them, and the corresponding data-collection procedure. 

\textbf{ATD lifecycle moments.}
Building on prior lifecycle-tracing work, we focus on three key moments:

\begin{itemize}
  \item \textbf{Introduction (source code):} the point where an architectural compromise is first introduced in the repository.
  \item \textbf{Detection (issue tracker):} the point where the ATD is documented/recognized in the Jira issue tracker.
  \item \textbf{Repayment (source code):} the point where the ATD is removed or mitigated through code changes.
\end{itemize}




For our lifecycle analysis, the interval of interest is the elapsed time between Introduction and Repayment, while Detection provides the issue-tracker context and serves as the entry point for identifying the ATD items.

\textbf{Operational definitions of the key moments.}
The \textbf{Introduction} and \textbf{Repayment} moments are observable in practice as Git commits. An introduction commit may add a workaround, hard-coded configuration, shortcut, or dependency structure that violates intended modular boundaries. A repayment commit removes or mitigates the shortcut by restoring architectural boundaries, introducing missing abstractions, or realigning dependencies with the intended design. The \textbf{Detection} moment corresponds to the Jira issue in which the architectural concern is documented. As established in our prior study~\cite{sutoyo2026reducinglabelingeffortarchitecture}, we manually examined issue reports for the ATD indicators VioMod and ObsTech. Two annotators independently assessed the issue content, with disagreements resolved through consensus and third-author adjudication when necessary. The resulting labels distinguish True-ATD, Weak-ATD, and Non-ATD. In this study, these ATD issues serve as starting points for tracing their corresponding repayment and introduction commits in version-control history.

\textbf{Data source.} We used the Jira issue dataset introduced in our previous ATD-detection study~\cite{sutoyo2026reducinglabelingeffortarchitecture} and further refined it for lifecycle analysis.\footnote{\url{https://github.com/edisutoyo/atd-jira-issues}} The source study selected large Java frameworks, libraries, and software systems that use Jira, based on data availability, domain diversity, and sustained development activity, and required more than 5,000 reported Jira issues. We collected project and issue information up to January 7, 2025, which defines the upper boundary of the issue dataset used here. We did not impose a fixed observation window in years; instead, we reconstructed each ATD lifecycle using the available repository history from its repayment commit back to its inferred introduction commit. The repository characteristics reported in \autoref{tab:projects}, including the number of contributors and commits, were measured up to the same cutoff date.



\begin{table*}[htpb!]
	\caption{Characteristics of the ten projects used in this study.}
	\scriptsize
	\label{tab:projects}
	\centering
    \resizebox{\linewidth}{!}{%
	\begin{tabular}{c l l r r r r r l}
		\toprule
		No & Project & Domain & SLOC & \#Reported Issues & \#Resolved Issues & Contributors & \#Commits & First Commit \\
		\midrule
		1  & Camel     & Integration framework      & 1,800k & 22,068 & 16,478 & 1,457 & 73,652 & 2007-03-19 \\
		2  & Spark     & Analytics engine           & 1,442k & 55,155 & 42,348 & 3,102 & 43,187 & 2010-03-29 \\
		3  & Kafka     & Stream-processing software & 1,012k & 19,283 & 11,403 & 1,515 & 14,664 & 2011-08-01 \\
		4  & ActiveMQ  & Message broker             & 423k   & 9,712  & 5,512  & 210   & 11,952 & 2005-12-12 \\
		5  & Cassandra & Database                   & 798k   & 20,648 & 16,967 & 650   & 30,374 & 2009-03-02 \\
		6  & Drill     & Query engine               & 890k   & 8,524  & 3,571  & 259   & 4,527  & 2012-09-03 \\
		7  & Geode     & Data management            & 1,350k & 10,457 & 1,125  & 251   & 11,257 & 2015-04-28 \\
		8  & Lucene    & Search engine library      & 882k   & 10,681 & 2,438  & 554   & 37,804 & 2001-09-11 \\
		9  & NetBeans  & IDE                        & 5,400k & 6,519  & 272    & 317   & 10,802 & 2017-09-11 \\
		10 & Solr      & Load balancer              & 705k   & 17,762 & 3,847  & 547   & 37,847 & 2001-09-11 \\
		\bottomrule
	\end{tabular}
    }
\end{table*}

To make this paper self-contained, we briefly summarize the construction and labeling procedure of the source dataset used in this study. The dataset spans ten large-scale Apache projects (\autoref{tab:projects}), covering a broad range of domains. In the source study, issues were randomly sampled using a 95\% confidence level and a 5\% margin of error, and two authors independently labeled 3,313 instances (each instance corresponds to one Jira issue); disagreements were resolved by consensus, with third-author adjudication when necessary. The resulting Cohen's kappa coefficient was $\kappa = 0.70$, indicating substantial agreement between the two annotators \cite{Cohens_kappa}.

For labeling ATD from Jira issue reports, we retain the annotation framework used in our previous study~\cite{sutoyo2026reducinglabelingeffortarchitecture}, which builds on the technical-debt classification frameworks of Alves et al.~\cite{alves2014towards} and Li et al.~\cite{li2023automatic}. Within that framework, we focus on the two indicators explicitly associated with architectural technical debt: \textit{Violation of Modularity (VioMod)} and \textit{Use of Obsolete Technology (ObsTech)}. VioMod captures architectural shortcuts that introduce undesired interdependencies among modules that should remain independent, whereas ObsTech captures the continued use of an architecturally significant technology that has become obsolete. 

These two indicators do not constitute an exhaustive taxonomy of ATD. Other manifestations, such as architectural erosion, architectural smells, cyclic dependencies, and broader structural trade-offs, may also be relevant. We focus on VioMod and ObsTech because they are the ATD-specific indicators operationalized and expert-validated in the source Jira dataset, allowing this lifecycle study to build on an established and reproducible classification.

Jira issue type, component, and test-related terminology were not used alone to determine ATD. Classification was based on the architectural concern expressed in the issue content according to the VioMod and ObsTech indicators. Thus, test-related issues could still be classified as ATD when the underlying concern involved architectural modularity or obsolete technology.

To reflect varying degrees of clarity in architectural debt expressions, the annotation distinguishes three labels: \textit{True-ATD}, \textit{Weak-ATD}, and \textit{Non-ATD}. \textit{True-ATD} refers to issue reports that explicitly and unambiguously describe architectural concerns, trade-offs, or structural design decisions that may compromise long-term maintainability, scalability, or performance and can be labeled as ATD by all annotators. When disagreements occurred (e.g., \enquote{ATD} vs.\ \enquote{Non-ATD}), a third annotator was consulted, and majority voting determined the final label. In contrast, \textit{Weak-ATD} captures borderline issues that hint at architectural concerns but lack sufficient context for definitive classification; these cases were initially marked as \enquote{Maybe} by one annotator, while the other labeled them as ATD. Weak-ATD thus represents a softer interpretation of architectural debt that remains useful for training or complementary analyses under less strict criteria. In the end, the dataset contains 1,100 ATD items, and while smaller than large-scale mining datasets, its size is consistent with prior work~\cite{wen2019exploring,al2024detecting,prenner2021making,tsantalis2018accurate}.

Importantly, the presence of a repository change is not used to determine whether an issue constitutes ATD. ATD classification is established independently from the Jira issue content using the VioMod and ObsTech indicators; repository traceability is applied only afterward to determine whether its introduction and repayment lifecycle can be reconstructed.

\begin{figure*}[htpb] 
    \centerline{\includegraphics[trim=0.1cm 0.1cm 0.1cm 0.1cm, clip, width=0.9\textwidth]{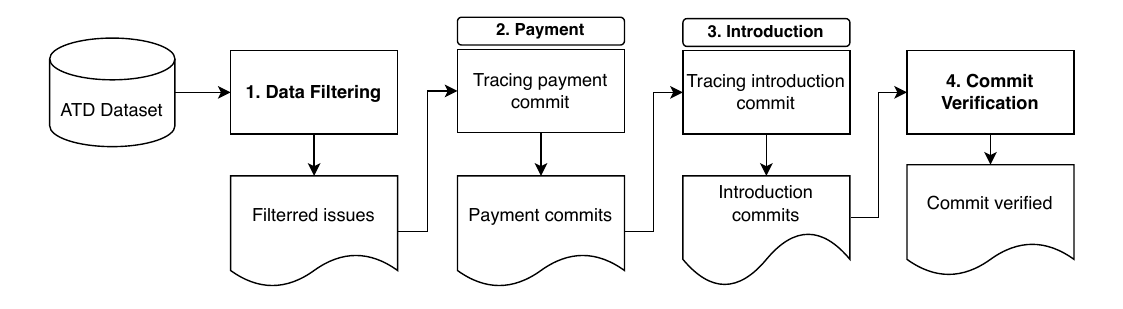}}
    \caption{Overview of the four-step data-collection and lifecycle-tracing procedure.} 
    \label{fig_lifecycle_steps}
\end{figure*}

\textbf{Data collection procedure.}
Starting from the detected ATD issues in Jira, we construct the ATD lifecycle dataset by (1) filtering issues, (2) mapping issues to repayment commits, and (3) tracing introduction commits from the repayment commits using an SZZ-style strategy implemented with PyDriller, followed by (4) sampled manual verification, as depicted in \autoref{fig_lifecycle_steps}.

\begin{enumerate}
    \item \textbf{Dataset filtering.} We remove invalid entries (e.g., duplicates) and restrict the lifecycle analysis to ATD issues whose repayment can be traced to a commit in the studied repository, as this is required to determine repayment time and fixer identity. Issues without an identifiable repayment commit are excluded because their lifecycle cannot be reconstructed using our Git-based approach, rather than because they are considered Non-ATD. Architectural debt may also be repaid through documentation, configuration, external services, operational processes, or artifacts outside the repository; such repayment modes are therefore outside the scope of this study.

    \item \textbf{Tracing repayment commits.} Since commit hashes are not directly stored in the dataset, we scan each repository's Git history and extract commits whose messages reference the Jira issue ID, as is common in Apache projects~\cite{claes202020}. We extract Jira identifiers from commit messages using the case-insensitive regular expression \texttt{\detokenize{(?i)\b<PREFIX>[\-\:_ ]?(\d{1,7})\b}}, where \texttt{PREFIX} denotes the project-specific Jira prefix. Each matched identifier is normalized to the canonical \texttt{PREFIX-NUMBER} format. We retain all Jira identifiers found in a commit message, allowing a single commit to be associated with multiple Jira issues. 

    Because the same Jira issue may be referenced by multiple commits, we consider all matching commits and select the chronologically latest mention as the candidate repayment commit. This reduces the likelihood of selecting an earlier intermediate implementation commit. Nevertheless, the presence of a Jira identifier in a commit message does not by itself guarantee that the selected commit corresponds to the definitive final repayment; we therefore treat this mapping as a repository-based approximation of the repayment event. 

    When a commit message references multiple Jira identifiers, the same commit may be selected as the candidate repayment commit for more than one issue. We do not partition the changed lines of such bulk commits among the individual Jira issues because Git metadata does not provide a reliable mapping between specific changed lines and the different issue identifiers. Consequently, the subsequent SZZ analysis is applied to the relevant changes in the complete selected commit for each associated ATD item. 
    
    Issues without any identifiable commit reference are excluded from the lifecycle analysis~\cite{tan2023lifecycle}.
    An explicit Jira issue reference provides strong issue-to-commit traceability, but does not guarantee that the linked commit captures the complete architectural repayment. We therefore treat it as the observable repository-level repayment point, while acknowledging that additional remediation may occur outside the repository. 

    \item \textbf{Tracing introduction commits.} We use \texttt{PyDriller}~\cite{PyDriller} to mine the repository history and recover candidate commits associated with lines modified during repayment. PyDriller also provides line-history tracing functionality that applies an SZZ-based procedure \cite{sliwerski2005changes}: given a commit, it identifies the changed lines and uses line-level blame to recover earlier commits that last modified those lines. Prior SATD studies have used PyDriller to extract code changes and trace SATD introduction and removal commits~\cite{bhatia2026empirical,muse2022fixme}.
    
    For each repayment commit, we use this functionality to recover candidate introduction commits from the lines deleted or modified during repayment. Because multiple earlier commits may be returned, we rank the candidates by the number of traced repayment-related lines attributed to each candidate and select the highest-ranked candidate as the inferred introduction commit; ties are resolved by selecting the earliest candidate.

    The candidate ranking is used only to select among the commits returned by the line-history tracing procedure and is not treated as independent evidence that the selected commit is necessarily the true ATD introduction point. Because ATD may span multiple files and artifacts, we therefore assess the reliability of the inferred lifecycle boundaries through a separate ATD-specific manual validation procedure described below.

    \item \textbf{Manual validation of ATD lifecycle tracing.} To assess whether the repository-based tracing procedure produces credible lifecycle boundaries for ATD, we manually inspected a random sample of 45 fully traced ATD items (approximately 5\% of the dataset).
    
    For each item, we cross-checked the Jira issue description, the repayment commit and its message and diff, and the inferred introduction commit. A repayment commit was confirmed only when the observed changes were consistent with the architectural concern and repayment described in Jira. An inferred introduction commit was confirmed only when the inspected changes provided evidence that the commit introduced code subsequently modified during repayment and that the change was consistent with the architectural concern described in the issue.
    
    All 45 repayment commits were confirmed (100\%). For the inferred introduction commits, 42 of 45 were confirmed, corresponding to a 93.3\% confirmation rate. The remaining three candidates were rejected because, although they modified related code, the available evidence was insufficient to establish that they introduced the architectural concern. 
    
\end{enumerate}

\autoref{tab:ds_label_distribution} reports the ATD label distribution before and after commit-tracking filtering. The initial dataset contained 1,100 ATD issues, comprising 690 \textit{True-ATD} and 410 \textit{Weak-ATD} issues across the VioMod and ObsTech categories. After restricting the dataset to issues that were successfully linked to their repayment commits, 896 ATD items remained. The remaining 204 issues had no identifiable Jira issue ID reference in any commit and were therefore excluded. For the main lifecycle analysis, we retain both True-ATD and Weak-ATD items. Although Weak-ATD items carry lower classification confidence, they are not classified as Non-ATD; rather, they contain indications of architectural concerns identified using the same VioMod and ObsTech annotation framework, but with less explicit supporting context. Retaining these items therefore allows the main analysis to capture a broader range of repository-observable ATD expressions while acknowledging uncertainty in borderline cases. To assess whether the inclusion of Weak-ATD items materially affects the findings, we additionally perform a robustness analysis restricted to the 577 True-ATD items; the complete analysis is provided in the replication package.\footnote{\url{https://github.com/edisutoyo/self-fixed-atd}}

\begin{table}[htb]
	\centering
	\caption{ATD label distribution before and after commit-tracking filtering.}
    \tiny
	\label{tab:ds_label_distribution}
	\begin{tabular}{lccc|ccc}
		\toprule
		& \multicolumn{3}{c}{Initial Dataset} 
		& \multicolumn{3}{c}{Final Dataset} \\
		\cmidrule(lr){2-4}
		\cmidrule(lr){5-7}
		Label 
		& VioMod & ObsTech & Count
		& VioMod & ObsTech & Count \\
		\midrule
		True-ATD
		& 532 & 158 & 690
		& 443 & 134 & 577 \\
		
		Weak-ATD
		& 334 & 76 & 410
		& 263 & 56 & 319 \\
		
		\midrule
		\textbf{Total}
		& \textbf{866} & \textbf{234} & \textbf{1,100}
		& \textbf{706} & \textbf{190} & \textbf{896} \\
		\bottomrule
	\end{tabular}
\end{table}

\subsection{Data Analysis}\label{subsecDataAnalysis}
Using the ATD lifecycle dataset constructed in the previous subsection (\autoref{fig_lifecycle_steps}), we performed the following analyses. For each ATD item, we identified the introducer (author of the introduction commit) and the fixer (author of the repayment commit) from version-control metadata. When the introducer and fixer are the same person, the item is categorized as \textbf{self-fixed ATD}; otherwise, it is categorized as \textbf{non--self-fixed ATD}.

\subsubsection{RQ1—Prevalence and who repays ATD.} To address \textbf{RQ1}, we quantified how frequently ATD items were self-fixed relative to all repaid items using the self-fixing rate:

\begin{equation}
\textit{self-fixing rate} = \frac{\textit{num\_of\_self-fixed}}{\textit{num\_of\_fixed}}
\label{eq:self-fixing-rate}
\end{equation}
where $\textit{num\_of\_self-fixed}$ is the number of self-fixed ATD items, and $\textit{num\_of\_fixed}$ is the total number of ATD items that have been remediated.

While RQ1 captures \emph{whether} the introducer ultimately repays the ATD, it does not capture how development activity is distributed across affected files over the item's lifetime. To address this, we analyze file-level contribution shares on ATD-affected files from the introduction commit through the repayment commit using two complementary measures: (i) commit frequency and (ii) change size. We define change size as the number of added plus deleted lines, capturing the magnitude of code modification.

Because the relevant contributor roles differ between self-fixed and non--self-fixed ATD, we characterize their contribution distributions separately. For non--self-fixed ATD, we distinguish the \textit{introducer}, the eventual \textit{fixer}, and \textit{Others} (all remaining developers). For self-fixed ATD, the introducer and fixer are the same developer; therefore, we distinguish the \textit{self-fixer} from \textit{Others}. The analysis comprises the following steps:

\begin{enumerate}
\item 

\textbf{Per-file contribution shares.}

For every (Key, File) pair, we compute role-specific contribution shares using commit count ($C$) and change size ($L$).

Let $X_{r,m}^{f}$ denote the contribution of role $r$ to file $f$ under
measure $m\in\{C,L\}$, where $X_{r,C}^{f}=C_{r}^{f}$ denotes the number
of commits and $X_{r,L}^{f}=L_{r}^{f}$ denotes the number of changed lines.
For non--self-fixed ATD,
$r\in\{\text{intro},\text{fix},\text{oth}\}$, whereas for self-fixed ATD,
$r\in\{\text{self},\text{oth}\}$.

For each file and contribution measure, the total contribution is the sum of contributions over the applicable developer-role set. The corresponding per-file contribution share is
\begin{equation}
\text{share}_{r,m}^{f}
=
\frac{X_{r,m}^{f}}
     {\sum_{r'} X_{r',m}^{f}},
\qquad
m\in\{C,L\},
\label{eq:per-file-contribution-share}
\end{equation}

where $r'$ ranges over
$\{\text{intro},\text{fix},\text{oth}\}$ for non--self-fixed ATD and
$\{\text{self},\text{oth}\}$ for self-fixed ATD.

To focus on files that were actively modified during the ATD item's lifetime, we retain only file--item pairs with at least one intermediate commit beyond the introduction and repayment commits. Change-size-based shares are computed for file--item pairs for which line-level change information can be reconstructed.

\item \textbf{Contribution-distribution analysis.}

For non--self-fixed ATD, we descriptively compare the distributions of the Introducer, eventual Fixer, and Others for both commit- and change-size-based contribution shares. For self-fixed ATD, we compare the Self-fixer and Others under both measures. This allows us to assess whether the observed contribution-concentration patterns are consistent across commit frequency and change magnitude.

\item \textbf{Hypothesis tests for self-fixed ATD.}

Because the Self-fixer and Others shares are complementary under both measures, the inferential question is whether the Self-fixer accounts for more than half of the contribution activity on an affected file. We therefore apply separate one-sample Wilcoxon signed-rank tests to $\mathrm{share}_{\mathrm{self},C}^{f}-0.5$ and $\mathrm{share}_{\mathrm{self},L}^{f}-0.5$. As a robustness check, we additionally apply a one-sided sign test against the 0.5 threshold.

For commit-based contributions, we test
$H_{1,C}:
\mathrm{median}(\mathrm{share}_{\mathrm{self},C}^{f})>0.5$.
For change-size-based contributions, we test
$H_{1,L}:
\mathrm{median}(\mathrm{share}_{\mathrm{self},L}^{f})>0.5$.
For both Wilcoxon tests, we report the rank-biserial correlation (RBC) \cite{cureton1956rank} as the corresponding effect size.

\item \textbf{Effect summaries.}

For both contribution measures, we report the median and interquartile range (IQR) of the Self-fixer share, the proportion of file--item pairs for which the Self-fixer share exceeds 0.5, and the RBC from the Wilcoxon test. For non--self-fixed ATD, we summarize and visualize the corresponding distributions for the Introducer, eventual Fixer, and Others.

\end{enumerate}

\subsubsection{RQ2—Time-to-fix and repayment dynamics.} To address \textbf{RQ2}, we estimate the survival time of each ATD item, defined as the number of days from its introduction to its remediation in the codebase:

\begin{equation}
\textit{TTF}(\textit{issue}) = \textit{date}_{\textit{payment}} - \textit{date}_{\textit{intro}}
\label{eq:ttf}
\end{equation}

We treat each ATD item as an observation with a repayment event at \textit{TTF} days. We compare self-fixed and non--self-fixed ATD using Kaplan--Meier (KM) cumulative-fix curves, reporting the complete observed repayment window as the primary analysis. Following the early-repayment window used by Tan et al.~\cite{tan2022does}, we additionally report a 30-day zoom to facilitate comparison with prior self-fixed TD research and to examine whether differences between repayment modes emerge early in the ATD lifecycle. The 30-day window is used only as a descriptive early-stage reference and is not intended to represent a project-independent repayment threshold or an expected duration for resolving ATD.

This distinction is particularly important for ObsTech items, whose repayment may require longer-term technology migration or replacement activities; therefore, the 30-day analysis for ObsTech should be interpreted only as an early-repayment comparison, while the full-window curves capture its longer-term repayment behavior.


Beyond \enquote{when} ATD is repaid, we examine \enquote{how} repayment unfolds via developer involvement on ATD-affected files. Because introducers may still contribute even when they are not the final repayment author, we analyze role-specific contributions for each non--self-fixed issue over the introduction--payment interval.

To address \textbf{RQ2.1}, we count commits from the \textit{introduction} commit through (and including) the \textit{payment} commit, aggregating across all ATD-affected files for the issue. Let $C_\text{intro}^{\text{ATD}}$, $C_\text{fix}^{\text{ATD}}$, and $C_\text{oth}^{\text{ATD}}$ denote the number of commits to ATD-related files authored by the introducer, the fixer, and all other developers, respectively. We quantify each role's relative involvement using an \emph{involvement ratio} (IR). For example, the \emph{Introducer Involvement Ratio} (IIR) is:

\begin{equation}
\label{eq:ir}
\text{Introducer Involvement Ratio (IIR)} =
\frac{C_\text{intro}^{\text{ATD}}}{C_\text{intro}^{\text{ATD}} + C_\text{fix}^{\text{ATD}} + C_\text{oth}^{\text{ATD}}}
\end{equation}

By changing the numerator, we define three measures: \emph{Introducer Involvement Ratio} (IIR), \emph{Fixer Involvement Ratio} (FIR), and \emph{Other-developers Involvement Ratio} (OIR). Each IR is the share of commits on ATD-related files authored by the corresponding role among all commits by the introducer, fixer, and others during the ATD lifetime. IR values lie in $[0,1]$, where $1$ indicates full concentration in that role and $0$ indicates no activity. 


For comparisons, we stratify non--self-fixed ATD items into Low/Mid/High groups using issue-level IR values and quartile cutoffs. For each role-specific IR, let $q_{25}$ and $q_{75}$ be the 25th and 75th percentiles (Q1/Q3). Based on these thresholds, we define:
\[
\text{Low IR}:\ \mathrm{IR}\le q_{25},\qquad
\text{Mid IR}:\ q_{25}<\mathrm{IR}<q_{75},\qquad
\text{High IR}:\ \mathrm{IR}\ge q_{75}.
\]

We compare the three involvement groups using a global log-rank test and a Kruskal--Wallis test, followed by pairwise log-rank tests with Bonferroni correction.


\subsubsection{RQ3—Development factors associated with repayment time and responsibility.} \label{subsubsectionRQ3}
To address \textbf{RQ3}, we examine (i) how development factors differ between self-fixed and non--self-fixed ATD and (ii) how these factors are associated with time-to-fix. As discussed in Section~\ref{secRelatedWork}, prior studies on software maintenance and self-fixed TD have examined code-change characteristics, development activity, developer involvement, and project-specific developer experience in relation to maintenance and repayment outcomes. In particular, Tan et al.~\cite{tan2022does} examined several development- and developer-related characteristics surrounding TD repayment. Building on this prior work, we operationalize these factors using repository-observable information available for ATD lifecycles. We consider two complementary metric families:

\begin{itemize}
  \item \textbf{Code change characteristics (change scope).} We operationalize change scope using two complementary dimensions: change size and change breadth. Using the changed-line measure defined for RQ1.1, we aggregate change size at the ATD-item level as the cumulative code churn on the tracked ATD files, i.e., the total number of lines added and deleted by commits modifying these files during the introduction--repayment interval. Change breadth is measured as the number of distinct files modified by those ATD-associated commits, where an associated commit is a commit that modifies at least one tracked ATD file. A file is counted only once regardless of how many associated commits modify it. Thus, change size captures the magnitude of modifications on the tracked ATD files, whereas change breadth captures how broadly the associated development activity extends across the codebase.
  \item \textbf{Development activity.} We measure development activity using the total number of commits and the number of distinct developers involved during the introduction--repayment interval. The former captures the amount of development activity, whereas the latter captures how broadly that activity is distributed across developers.
\end{itemize}

These metrics represent repository-observable dimensions of change scope and development activity and are not intended to capture all technical, organizational, or human factors associated with ATD repayment.

We compare metric distributions between self-fixed and non--self-fixed items using the Mann--Whitney U test \cite{mann1947test} and report Cliff's $\delta$ \cite{cliff1993dominance} as an effect size. To assess whether a larger scope or more dispersed activity is associated with longer repayment, we compute Spearman's rank correlation coefficient $\rho$ \cite{spearman1961proof} between time-to-fix and each development factor.

Beyond these development-process factors, we additionally examine project-specific developer experience. Following Tan et al.~\cite{tan2022does} and prior repository-based studies of developer experience~\cite{eyolfson2014correlations}, we use project seniority as a repository-observable proxy for experience within the project. We define project seniority as the elapsed time between a developer's first recorded commit in the project and the ATD event under analysis. Accordingly, we measure introducer seniority at the introduction date, whereas we measure fixer seniority at the repayment date.

To further examine how development activity is distributed across developer roles in non--self-fixed ATD, we aggregate changed lines and commits separately for the Introducer, eventual Fixer, and Others. Because these role-specific measurements are repeated within the same ATD item, we use the Friedman test \cite{friedman1937use} as the global omnibus test, followed by paired Wilcoxon signed-rank tests with Holm correction \cite{wilcoxon1945individual}. We report RBC \cite{cureton1956rank} as the effect size for the pairwise comparisons.

We emphasize that this measure captures project-specific tenure rather than a developer's overall professional experience or architectural expertise. Therefore, seniority is not treated as the sole developer-related factor. In our multivariate analysis, we complement it with the introducer's contribution activity and breadth, represented by the total number of commits and the total number of distinct files touched in the project. Together, these repository-derived measures provide complementary information about how long, how actively, and how broadly a developer has contributed to the project, while acknowledging that characteristics such as professional experience outside the project, architectural role, and domain expertise cannot be observed reliably from repository data.

We compare seniority distributions between self-fixed and non--self-fixed items using Mann--Whitney U tests with Cliff's $\delta$. To jointly examine the likelihood that an ATD item is self-fixed while accounting for multiple developer-related characteristics and project- and developer-level heterogeneity, we fit a logistic generalized linear mixed model (GLMM) \cite{mcculloch2003generalized}. The binary response indicates whether an ATD item is self-fixed (1) or non--self-fixed (0). As fixed effects, we include the introducer's seniority at introduction, total number of commits, and total number of distinct files touched in the project, all standardized as z-scores. We include random intercepts for projects and introducers and fit the model using a binomial logit link with the \texttt{lme4} package in \texttt{R}.

\section{Results}\label{secResults}

\subsection{RQ1—Prevalence of self-fixing in ATD repayment}\label{subsecrq1}
To assess the prevalence of self-fixing, we examined whether the developer who introduced an ATD item was also responsible for its repayment. Overall, 166 of the 896 traceable ATD items were self-fixed, resulting in a self-fixing rate of 18.5\%. Using the metric defined in the methodology (see \autoref{eq:self-fixing-rate}), we compute:

\[ 
\textit{self-fixing rate} = \frac{166}{896} = 0.185 \approx 18.5\%. 
\]

Thus, the majority of the analyzed ATD items (730; 81.5\%) were repaid by developers other than their original introducers. This result indicates that, within the analyzed ATD items, repayment was predominantly performed by developers other than those who originally introduced the debt.

To assess whether self-fixing differs by the type of architectural concern, we further compute the self-fixing rates for the two ATD indicators, VioMod and ObsTech:

\[
\begin{aligned} \textit{self-fixing rate}_{VioMod} &= \frac{133}{706} = 0.188 \approx 18.8\%,\\
\textit{self-fixing rate}_{ObsTech} &= \frac{33}{190} = 0.174 \approx 17.4\%. \end{aligned} 
\]

\begin{figure*}[htpb]
\centerline{ \includegraphics[ trim=0.1cm 0.1cm 0.1cm 0.1cm, clip, width=0.4\textwidth ]{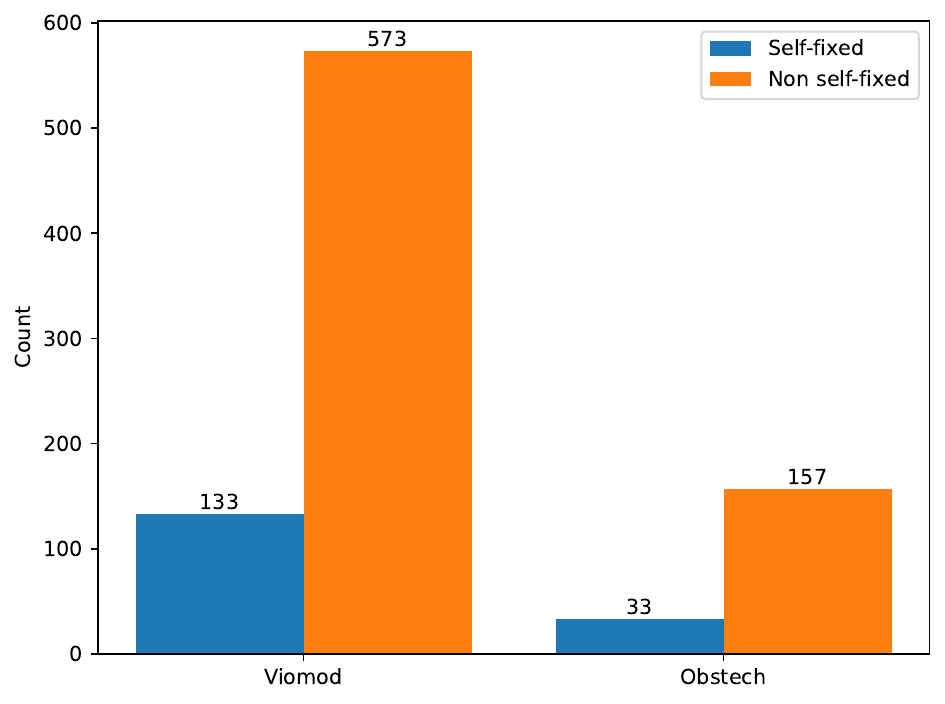} }
\caption{Distribution of self-fixed and non--self-fixed ATD items grouped by indicator.} 
\label{fig_self_fix_by_indicator} 
\end{figure*}

As shown in \autoref{fig_self_fix_by_indicator}, the two indicators exhibit similar observed self-fixing proportions. Specifically, 133 of the 706 VioMod items (18.8\%) and 33 of the 190 ObsTech items (17.4\%) were repaid by the same developers who introduced them. Although VioMod accounts for a substantially larger number of analyzed ATD items, self-fixing remains relatively uncommon for both architectural concern categories.

To explore whether the observed self-fixing pattern differs across projects, we additionally report the project-level distribution of the analyzed ATD items in \autoref{tab:self-fixed-per-project}. These values are descriptive statistics for the ATD items included in our dataset rather than estimates of the complete population of ATD items within each project, since the dataset contains only a subset of issues from the ten studied projects.

\begin{table}[htbp]
\centering
\tiny
\caption{Observed self-fixed and non--self-fixed ATD proportions across the analyzed project samples.}
\label{tab:self-fixed-per-project}
\begin{tabular}{lrrr}
\toprule
Project & Self-Fixed (\%) & Non--Self-Fixed (\%) & Analyzed ATD Items \\
\midrule
Camel     & 21.43 & 78.57  & 126 \\
Spark     & 25.74 & 74.26  & 101 \\
Kafka     & 10.86 & 89.14  & 175 \\
ActiveMQ  & 13.04 & 86.96  & 46  \\
Cassandra & 21.66 & 78.34  & 217 \\
Drill     & 12.82 & 87.18  & 39  \\
Geode     & 0.00  & 100.00 & 2   \\
Lucene    & 26.51 & 73.49  & 83  \\
NetBeans  & 25.00 & 75.00  & 4   \\
Solr      & 12.62 & 87.38  & 103 \\

\midrule
\textbf{Overall}
          & \textbf{18.53}
          & \textbf{81.47}
          & \textbf{896} \\
\bottomrule
\end{tabular}
\end{table}

Within the analyzed samples, the observed proportion of self-fixed ATD varies across projects. Among projects represented by larger numbers of analyzed ATD items, Lucene and Spark exhibit relatively high observed self-fixing proportions of 26.51\% and 25.74\%, respectively. In contrast, Kafka, Solr, and Drill show lower proportions of 10.86\%, 12.62\%, and 12.82\%, respectively. Camel and Cassandra exhibit intermediate and similar rates of 21.43\% and 21.66\%. The percentages for Geode and NetBeans should be interpreted with particular caution because these projects contain only two and four analyzed ATD items, respectively. Consequently, a single observation can substantially affect their project-level percentages.

Overall, self-fixing remains relatively uncommon across both ATD indicators and the analyzed projects. Although the observed rates vary across projects, these differences should be interpreted descriptively because the analyzed ATD items do not represent the complete ATD population of each project.





\begin{tcolorbox}[colback=gray!5!white, colframe=black, title=Summary of Findings (RQ1), boxrule=0.01pt, fonttitle=\small, fontupper=\small]
Self-fixing is uncommon in ATD: only 18.5\% of items are repaid by their introducers, with similarly low rates for VioMod (18.8\%) and ObsTech (17.4\%).  
\end{tcolorbox}


\subsection{RQ1.1—Distribution of contribution shares on ATD-affected files} \label{subsecrq1.1}
Beyond examining how frequently ATD items are self-fixed, we investigate how developer contributions are distributed across ATD-affected files for both self-fixed and non--self-fixed ATD. We analyze contribution at the file level using two complementary measures: the proportion of commits and the proportion of changed lines (change size). To focus on files with development activity during the ATD lifetime, we restrict the analysis to file--item pairs that receive additional commits between the introduction and repayment commits.

\begin{figure*}[htpb] 
    \centering

    \begin{minipage}[t]{0.4\textwidth}
        \centering
        \includegraphics[
            trim=0.1cm 0.1cm 0.1cm 0.1cm,
            clip,
            width=\textwidth
        ]{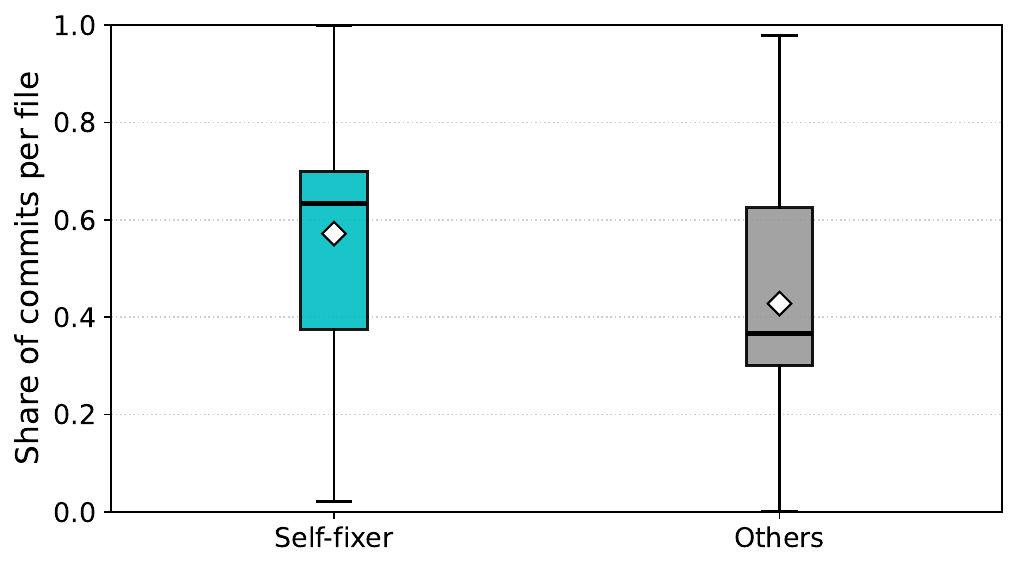} 
        \caption*{\textcolor{blue}{(a) Commit contribution share}}
    \end{minipage}
    \hfill
    \begin{minipage}[t]{0.4\textwidth}
        \centering
        \includegraphics[
            trim=0.1cm 0.1cm 0.1cm 0.1cm,
            clip,
            width=\textwidth
        ]{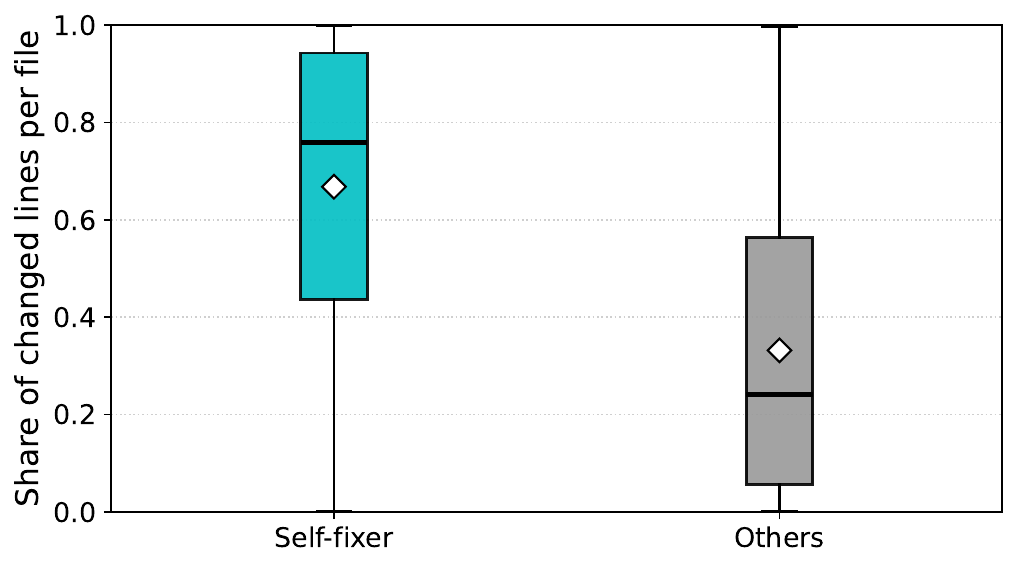} 
        \caption*{\textcolor{blue}{(b) Change-size contribution share}}
    \end{minipage}

    \caption{\textcolor{blue}{Boxplots of per-file contribution shares for the Self-fixer and Others in self-fixed ATD, based on commit counts and change size.}}
    \label{fig_rq2_box_SELF-FIXER_vs_OTHERS_WITH_ENDPOINTS}
\end{figure*}

\autoref{fig_rq2_box_SELF-FIXER_vs_OTHERS_WITH_ENDPOINTS} presents the contribution-share distributions for self-fixed ATD, contrasting the self-fixer with all remaining developers (Others). Because $\text{share}_{\text{oth}}^{f}=1-\text{share}_{\text{self}}^{f}$
for both measures (see Subsection~\ref{subsecDataAnalysis}), we examine whether the self-fixer accounts for more than half of the contribution activity on an affected file. This allows us to assess whether self-fixed repayment is associated with concentrated contribution by the self-fixer in both commit frequency and the magnitude of code changes.

For commit-based contributions, the median self-fixer share is 0.6337 ($Q_1=0.3750$, $Q_3=0.7000$), with 55.0\% of file--item pairs exceeding the 0.5 majority threshold. A one-sample Wilcoxon signed-rank test on $(\text{share}_{\text{self}}^{f}-0.5)$ confirms that the self-fixer share is significantly greater than this threshold ($W=4.88\times10^{6}$, $p=2.47\times10^{-49}$), consistent with the sign test ($p=1.76\times10^{-52}$). The effect size (RBC $=0.2705$) indicates a moderate tendency toward self-fixer dominance.

The pattern is stronger when contribution is measured by changed lines. Among file--item pairs for which line-level change information could be reconstructed, the median self-fixer share increases to 0.7589 ($Q_1=0.4369$, $Q_3=0.9430$), and 69.8\% exceed the 0.5 threshold. The Wilcoxon test again shows a significant positive deviation from 0.5 ($W=249{,}983$, $p=8.06\times10^{-45}$), supported by the sign test ($p=1.63\times10^{-30}$). The larger effect size (RBC $=0.5722$) indicates substantially stronger self-fixer dominance in change magnitude than in commit frequency.

\begin{figure*}[htpb] %
    \centering

    \begin{minipage}[t]{0.4\textwidth}
        \centering
        \includegraphics[
            trim=0.1cm 0.1cm 0.1cm 0.1cm,
            clip,
            width=\textwidth
        ]{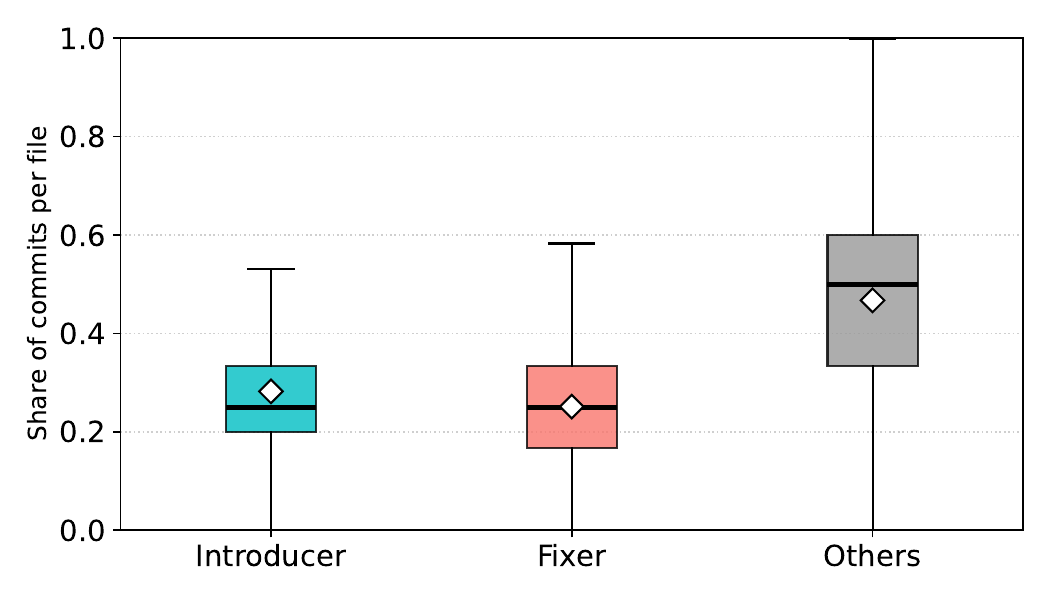} 
        \caption*{\textcolor{blue}{(a) Commit contribution share}}
    \end{minipage}
    \hfill
    \begin{minipage}[t]{0.4\textwidth}
        \centering
        \includegraphics[
            trim=0.1cm 0.1cm 0.1cm 0.1cm,
            clip,
            width=\textwidth
        ]{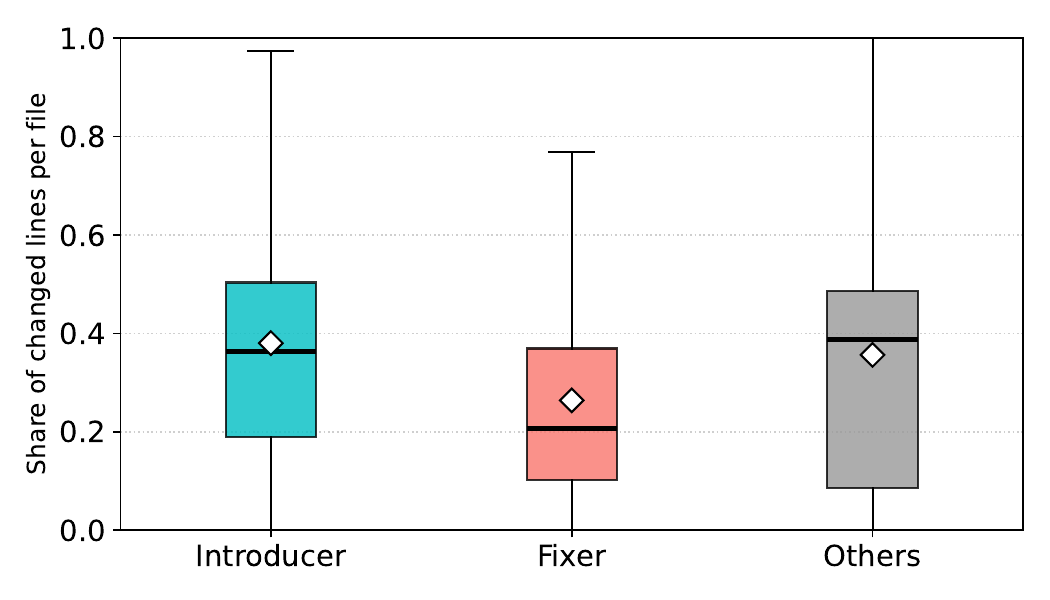} 
        \caption*{\textcolor{blue}{(b) Change-size contribution share}}
    \end{minipage}

    \caption{\textcolor{blue}{Boxplots of per-file contribution shares for the Introducer, Fixer, and Others in non--self-fixed ATD.}}
    \label{fig_rq2_box_RATIO_intro_vs_pay_per_file_WITH_ENDPOINTS_INTRO-FIXER-OTHERS}
\end{figure*}

For non--self-fixed ATD, \autoref{fig_rq2_box_RATIO_intro_vs_pay_per_file_WITH_ENDPOINTS_INTRO-FIXER-OTHERS}
compares the introducer, the eventual fixer, and the remaining developers (Others). For commit-based contributions, Others generally account for a larger per-file share than either the introducer or the eventual fixer, while the introducer and fixer show comparatively similar distributions. When contribution is measured by changed lines, the distributions become more dispersed: both the introducer and Others exhibit comparatively larger shares than the eventual fixer, with substantial variation across files.

Overall, the two measures reveal a clear difference between self-fixed and non--self-fixed ATD. Self-fixed ATD shows comparatively concentrated contribution by the self-fixer, particularly in terms of changed lines. In contrast, contribution activity in non--self-fixed ATD is more distributed across developer roles, and the eventual fixer generally accounts for a smaller share under both measures. Thus, the distinction between self-fixed and non--self-fixed ATD is reflected not only in who repays the debt, but also in how contribution activity is distributed over the affected files during the ATD lifetime.

\begin{tcolorbox}[colback=gray!5!white, colframe=black, title=Summary of Findings (RQ1.1), boxrule=0.01pt, fonttitle=\small, fontupper=\small]
Self-fixed ATD exhibits more concentrated contribution activity around the self-fixer, particularly in terms of changed lines, whereas non--self-fixed ATD shows more distributed contributions across developers.
\end{tcolorbox}



\subsection{RQ2—Time-to-fix comparison between self-fixed and non--self-fixed ATD}\label{subsecrq2}

\begin{figure*}[htpb] 
    \centerline{\includegraphics[trim=0.1cm 0.1cm 0.1cm 0.1cm, clip, width=0.55\textwidth]{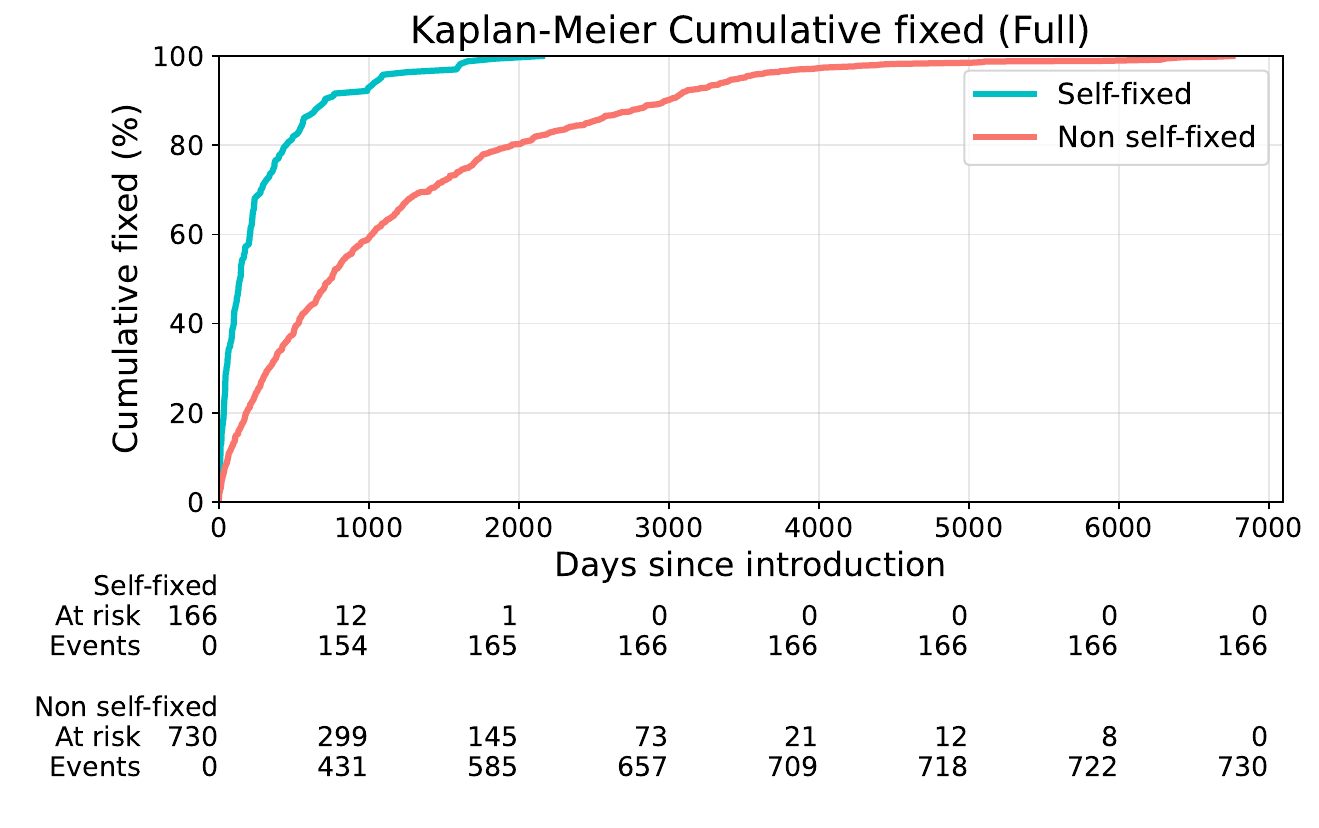}}
    \caption{Kaplan–Meier cumulative fixed curves comparing self-fixed and non--self-fixed ATD items across the entire observation period.} 
    \label{fig_KM_RQ2_full}
 \end{figure*}

\begin{figure*}[htpb] 
    \centerline{\includegraphics[trim=0.1cm 0.1cm 0.1cm 0.1cm, clip, width=0.55\textwidth]{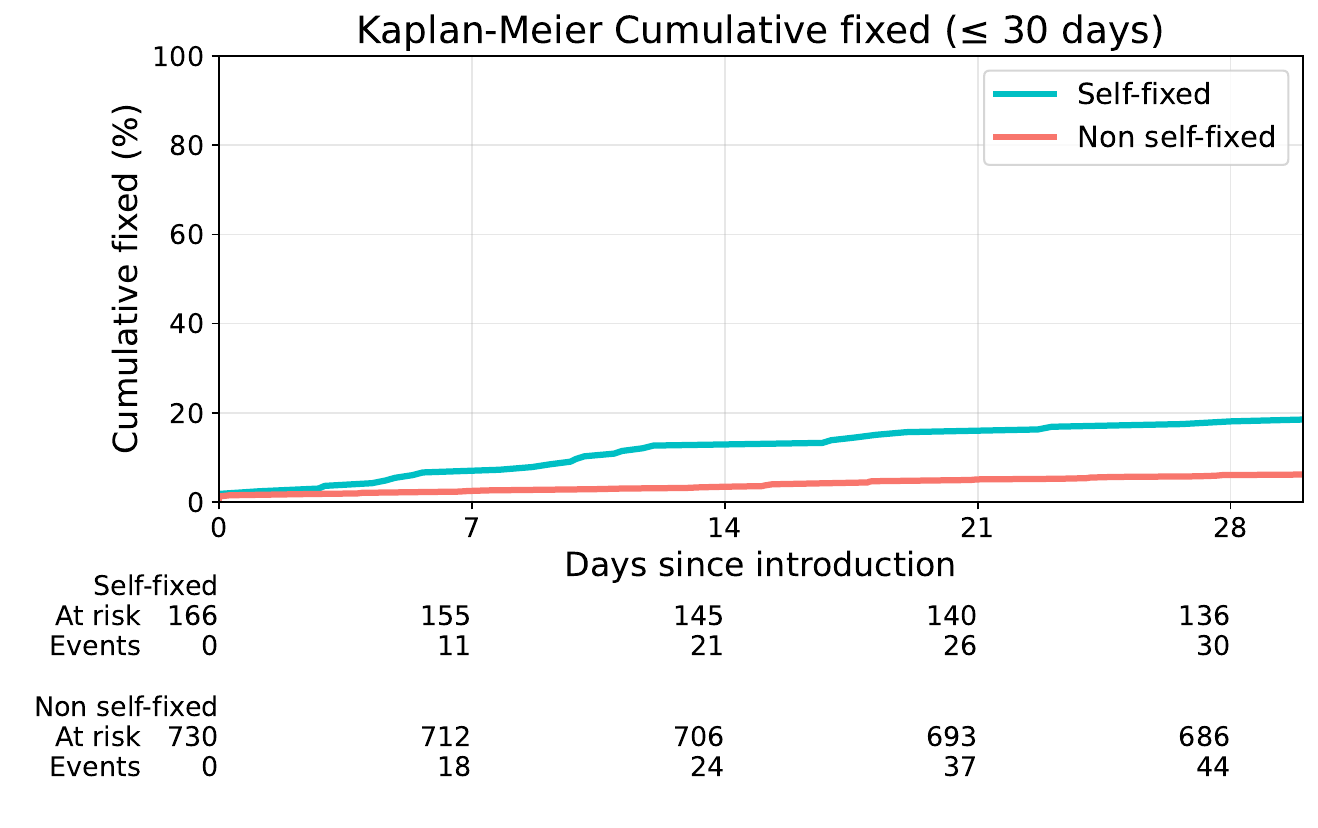}}
    \caption{Kaplan–Meier cumulative fixed curves for self-fixed and non--self-fixed ATD items within the first 30 days since introduction.} 
    \label{fig_KM_RQ2_30days}
 \end{figure*}

To answer RQ2, we used the non-parametric Kaplan--Meier estimator \cite{kaplan1958nonparametric} to compare the time-to-fix distributions of self-fixed ATD items and those fixed by other developers. \autoref{fig_KM_RQ2_full} depicts the cumulative proportion of repaid ATD items across the entire observed timeline, while \autoref{fig_KM_RQ2_30days} shows a zoomed-in perspective limited to the first 30 days after introduction. In these Kaplan--Meier plots, the x-axis denotes the number of days between introduction and repayment, and the y-axis reports the cumulative percentage of ATD items that have been repaid by each time point. For each time value, the numbers shown below the curves indicate the number of ATD items remaining unresolved, providing context for how many observations contribute to the estimated cumulative repayment at that point.

Across the full observation window (\autoref{fig_KM_RQ2_full}), self-fixed ATD items descriptively exhibit shorter time-to-fix. Approximately 90\% of self-fixed items are repaid within the first 1,000 days, while achieving a similar repayment proportion for non--self-fixed items requires more than 3,000 days. By the end of the observed period, self-fixed items reach full remediation slightly earlier than non--self-fixed items, indicating a descriptive association between self-fixing and shorter repayment time.

The zoomed-in view of the first 30 days (\autoref{fig_KM_RQ2_30days}) further highlights this early-stage difference. By day 30, approximately 18\% of self-fixed ATD has been repaid, compared to only 6\% of non--self-fixed ATD items. This early separation indicates that self-fixed ATD items exhibit earlier repayment than non--self-fixed items within the first 30 days.

\subsubsection{Survival time across indicator categories}
\begin{figure*}[htpb] 
    \centerline{\includegraphics[trim=0.1cm 0.1cm 0.1cm 0.1cm, clip, width=0.55\textwidth]{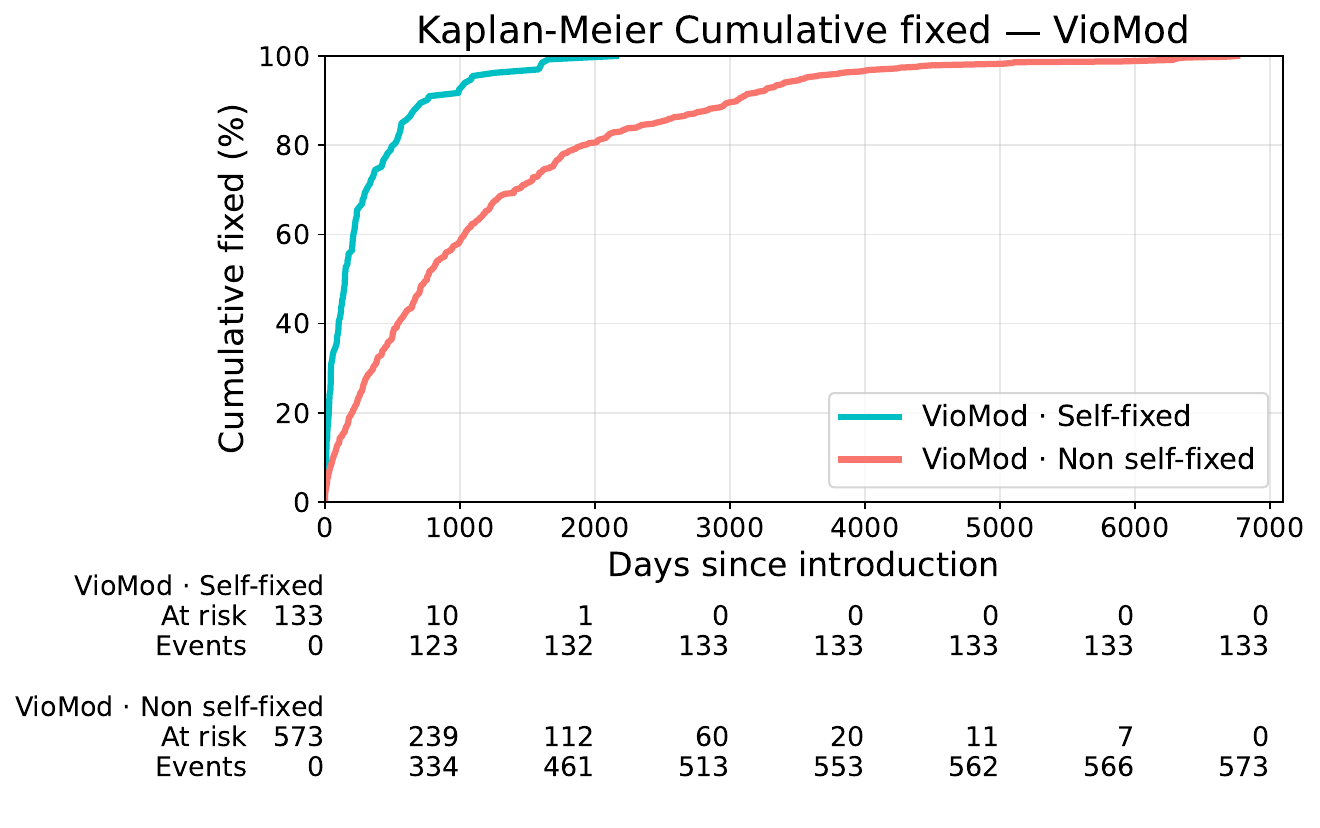}}
    \caption{Kaplan-Meier cumulative fixed curves for VioMod-related ATD items (full dataset).} 
    \label{fig_KM_VIOMOD_self_vs_nonself_full}
 \end{figure*}

\begin{figure*}[htpb] 
    \centerline{\includegraphics[trim=0.1cm 0.1cm 0.1cm 0.1cm, clip, width=0.55\textwidth]{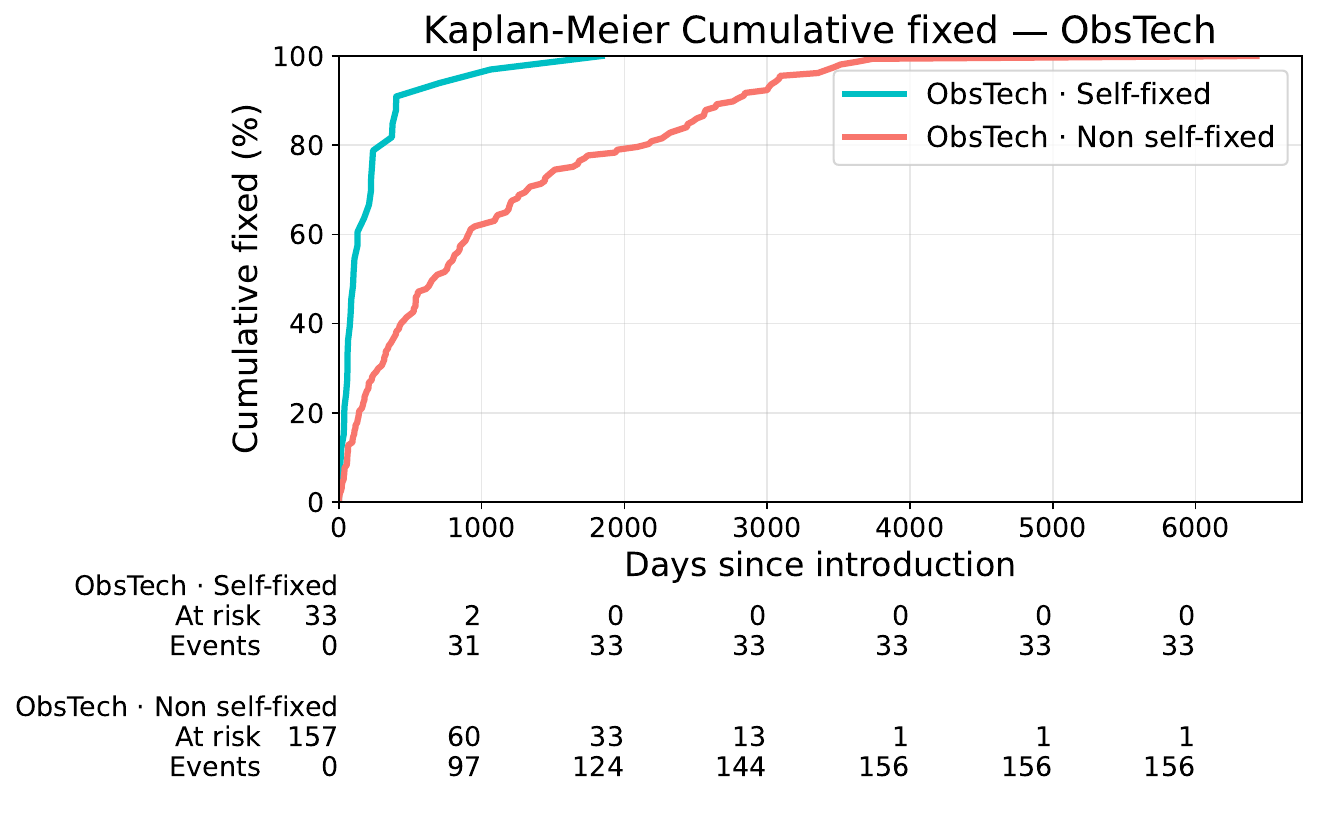}}
    \caption{Kaplan-Meier cumulative fixed curves for ObsTech-related ATD items (full dataset).} 
    \label{fig_KM_OBSTECH_self_vs_nonself_full}
 \end{figure*}

To examine whether repayment patterns differ across architectural-debt indicator categories, we analyze the survival time of ATD items for VioMod and ObsTech separately. The survival curves separate early in the timeline, indicating a stronger association between self-fixing and earlier repayment for VioMod than for non--self-fixed VioMod items. In each category, we compared ATD items that were self-fixed with the non--self-fixed ones. \autoref{fig_KM_VIOMOD_self_vs_nonself_full} and \autoref{fig_KM_OBSTECH_self_vs_nonself_full} present the Kaplan-Meier cumulative fixed plots based on the full dataset.

For the \textit{VioMod} category, self-fixed ATD items are resolved substantially faster than non--self-fixed items. More than 80\% of the self-fixed VioMod items are repaid within approximately 600-800 days, while non--self-fixed items reach the same repayment rate only after around 2,000-3,000 days. The survival curves clearly separate early in the timeline, indicating that when the original developer is responsible for remediation, modularity-related debt tends to be addressed much earlier. 

For ObsTech, the separation between self-fixed and non--self-fixed items is less pronounced than for VioMod. Self-fixed items show faster repayment earlier in the observation period, while the cumulative-fix curves gradually converge over time. A similar pattern is observed for ObsTech, although the separation between self-fixed and non--self-fixed items is less pronounced than for VioMod. Self-fixed ObsTech items show faster repayment during the earlier part of the observation period, while the cumulative-fix curves gradually converge over time. This indicates that the association between self-fixing and repayment time is weaker for ObsTech than for VioMod.

\begin{figure*}[htpb] 
    \centerline{\includegraphics[trim=0.1cm 0.1cm 0.1cm 0.1cm, clip, width=0.55\textwidth]{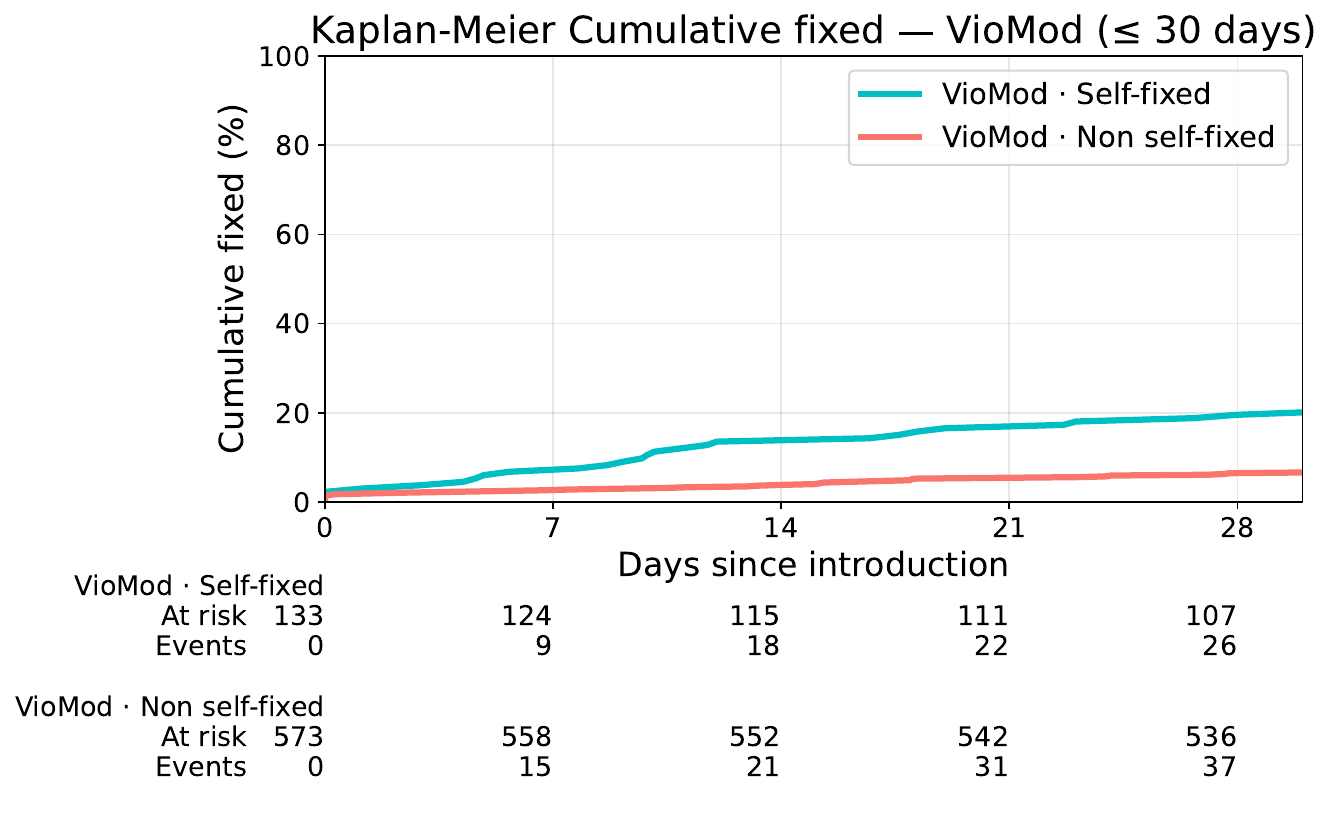}}
    \caption{Early 30-day Kaplan-Meier cumulative fixed for VioMod.}
    \label{fig_KM_VIOMOD_self_vs_nonself_30days}
 \end{figure*}

\begin{figure*}[htpb] 
    \centerline{\includegraphics[trim=0.1cm 0.1cm 0.1cm 0.1cm, clip, width=0.55\textwidth]{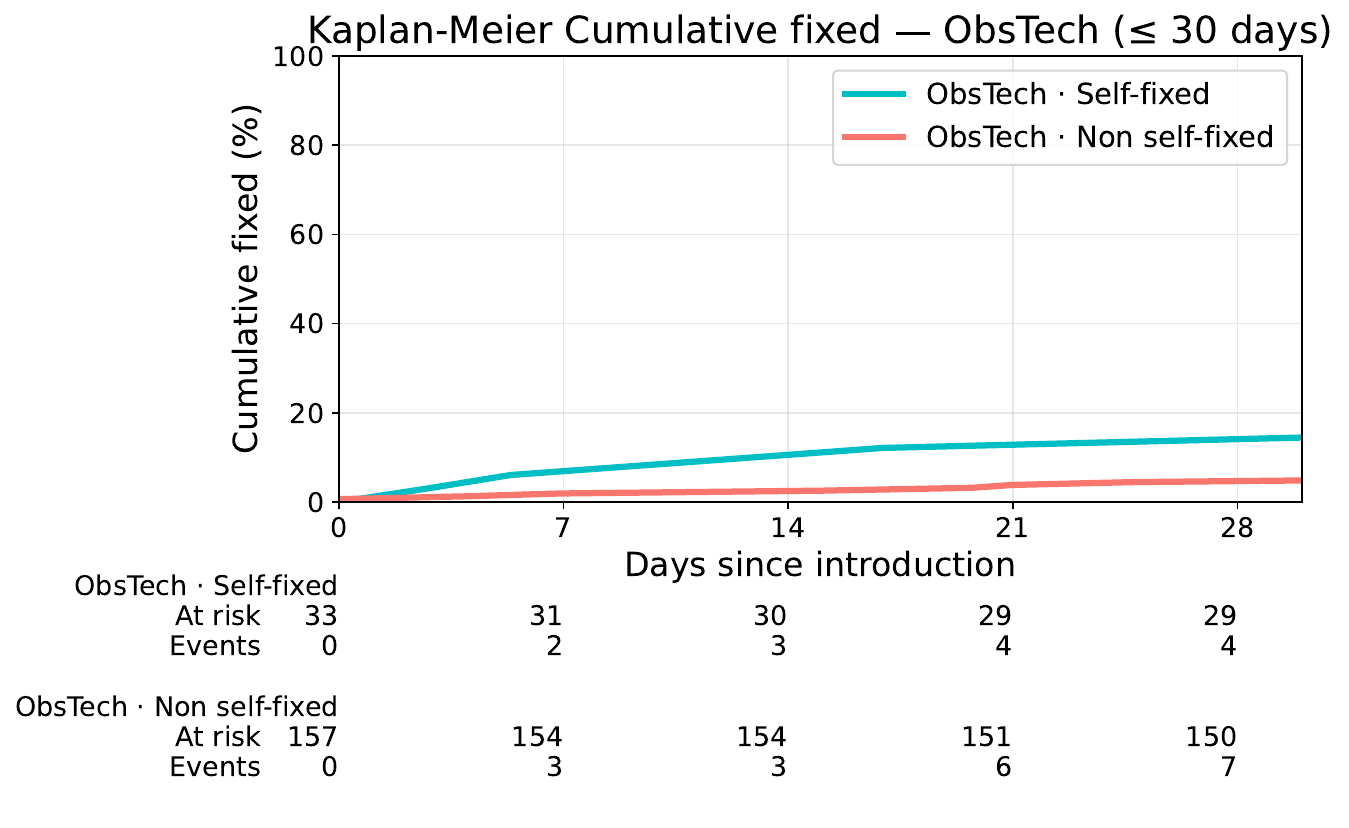}}
    \caption{Early 30-day Kaplan-Meier cumulative fixed for ObsTech.}
    \label{fig_KM_OBSTECH_self_vs_nonself_30days}
 \end{figure*}

Focusing on the first month after introduction, \autoref{fig_KM_VIOMOD_self_vs_nonself_30days} and \ref{fig_KM_OBSTECH_self_vs_nonself_30days} show that the self-fixed advantage appears within the first week and remains visible throughout the 30-day window, but it is stronger for VioMod than for ObsTech. For VioMod, the cumulative-fixed curve for self-fixed items rises steadily and reaches roughly $\approx$20\% by day 30, whereas non--self-fixed items remain around $\approx$6--7\%. For ObsTech, the separation is smaller: the self-fixed curve reaches only $\approx$14--15\% by day 30, while the non--self-fixed curve stays near $\approx$5--6\%, indicating a more modest early gap than in VioMod.

The 30-day view should be interpreted as an early-stage comparison rather than as a representative repayment horizon for either indicator. In particular, the smaller early separation for ObsTech does not imply that 30 days is sufficient for its repayment; the full-window analysis is used to characterize its longer-term repayment dynamics.

\begin{tcolorbox}[colback=gray!5!white, colframe=black, title=Summary of Findings (RQ2), boxrule=0.01pt, fonttitle=\small, fontupper=\small]
Self-fixed ATD is repaid substantially faster than non--self-fixed ATD: about 90\% of self-fixed items are repaid within 1{,}000 days, whereas non--self-fixed items require more than 3{,}000 days to reach a similar level. This pattern is consistent for both VioMod- and ObsTech-related ATD, where the self-fixed curves rise much earlier than the non--self-fixed curves. This gap highlights a practical danger, as non--self-fixing leaves projects exposed to unresolved ATD for much longer periods.
\end{tcolorbox}

\subsection{RQ2.1—Non--Self-fixed ATD and Developers' Involvement}
To better capture how work unfolds in practice, we consider that changes on ATD-affected files can be distributed across the introducer, the fixer, and other developers. Such activity may accumulate over the introduction--payment interval, and the introducer may contribute substantially even when the final repayment commit is authored by someone else. To assess whether different involvement patterns are associated with the time-to-fix of non--self-fixed ATD items, we compute a role-specific \emph{Involvement Ratio} (IR) for each issue (see~\ref{subsecDataAnalysis}).

\begin{figure*}[htpb]
    \centering

    \begin{minipage}[t]{0.3\textwidth}
        \centering
        \includegraphics[
            trim=0cm 0cm 0cm 0cm,
            clip,
            width=\textwidth
        ]{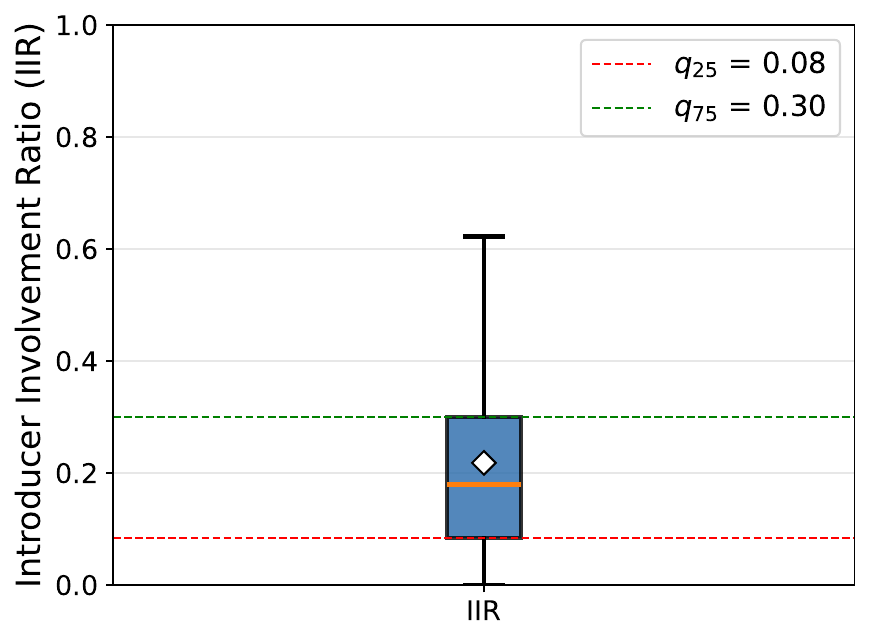}
        \caption*{(a) IIR}
    \end{minipage}\hfill
    \begin{minipage}[t]{0.3\textwidth}
        \centering
        \includegraphics[
            trim=0cm 0cm 0cm 0cm,
            clip,
            width=\textwidth
        ]{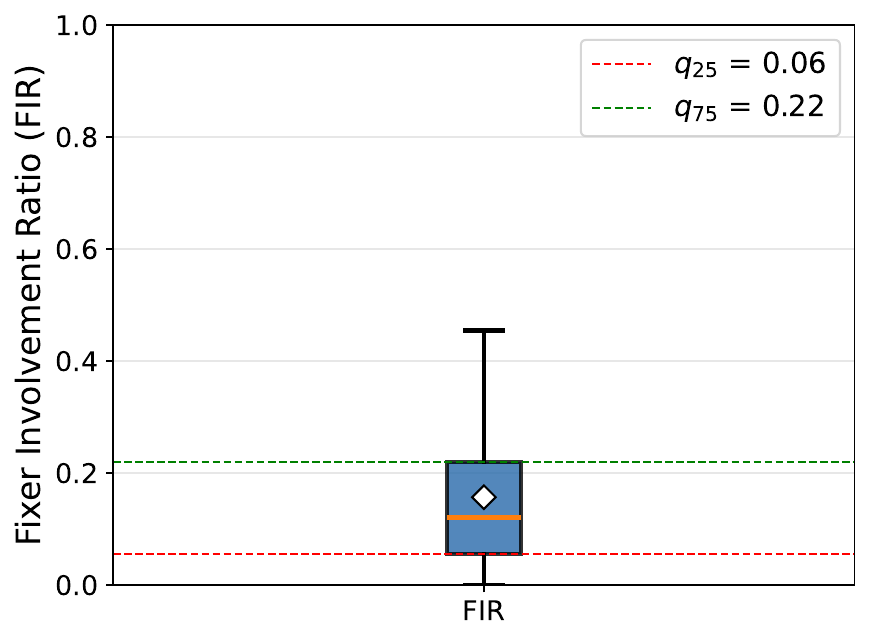}
        \caption*{(b) FIR}
    \end{minipage}\hfill
    \begin{minipage}[t]{0.3\textwidth}
        \centering
        \includegraphics[
            trim=0cm 0cm 0cm 0cm,
            clip,
            width=\textwidth
        ]{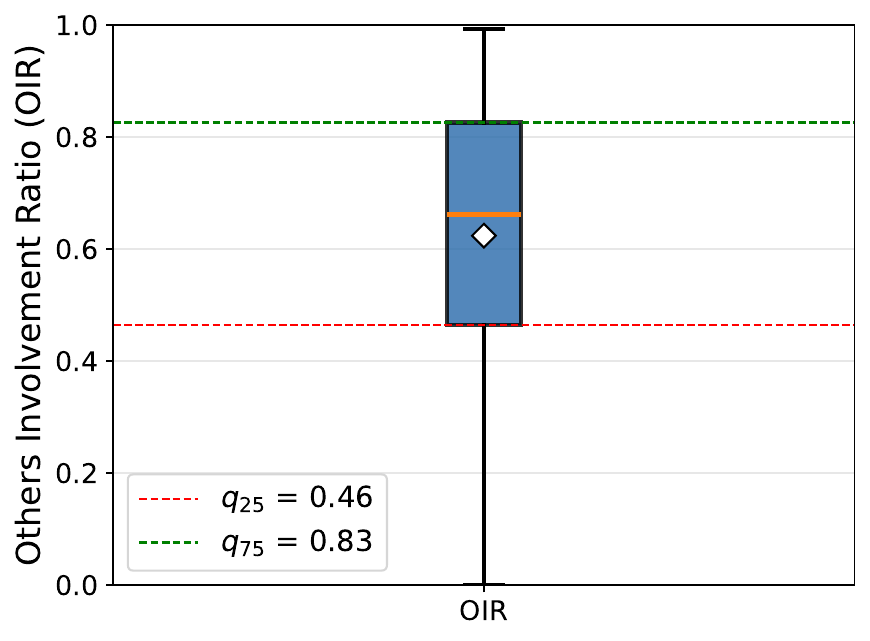}
        \caption*{(c) OIR}
    \end{minipage}

    \caption{Boxplots of the role-specific involvement ratios for non--self-fixed ATD items, showing the empirical $q_{25}$ and $q_{75}$ thresholds used to define the low-, mid-, and high-involvement groups.}
    \label{fig:role_ir_boxplots}
\end{figure*}

\autoref{fig:role_ir_boxplots} shows the observed distributions of IIR, FIR, and OIR for non--self-fixed ATD items, together with their corresponding quartile cutoffs. In our dataset, the empirical quartiles are $q_{25}=0.08$ and $q_{75}=0.30$ for IIR, $q_{25}=0.06$ and $q_{75}=0.22$ for FIR, and $q_{25}=0.46$ and $q_{75}=0.83$ for OIR. These thresholds define the low-, med-, and high-involvement groups used
in the subsequent survival analyses.

\begin{figure*}[htpb!] 
    \centerline{\includegraphics[trim=0.0cm 0.0cm 0.0cm 0.0cm, clip, width=0.55\textwidth]{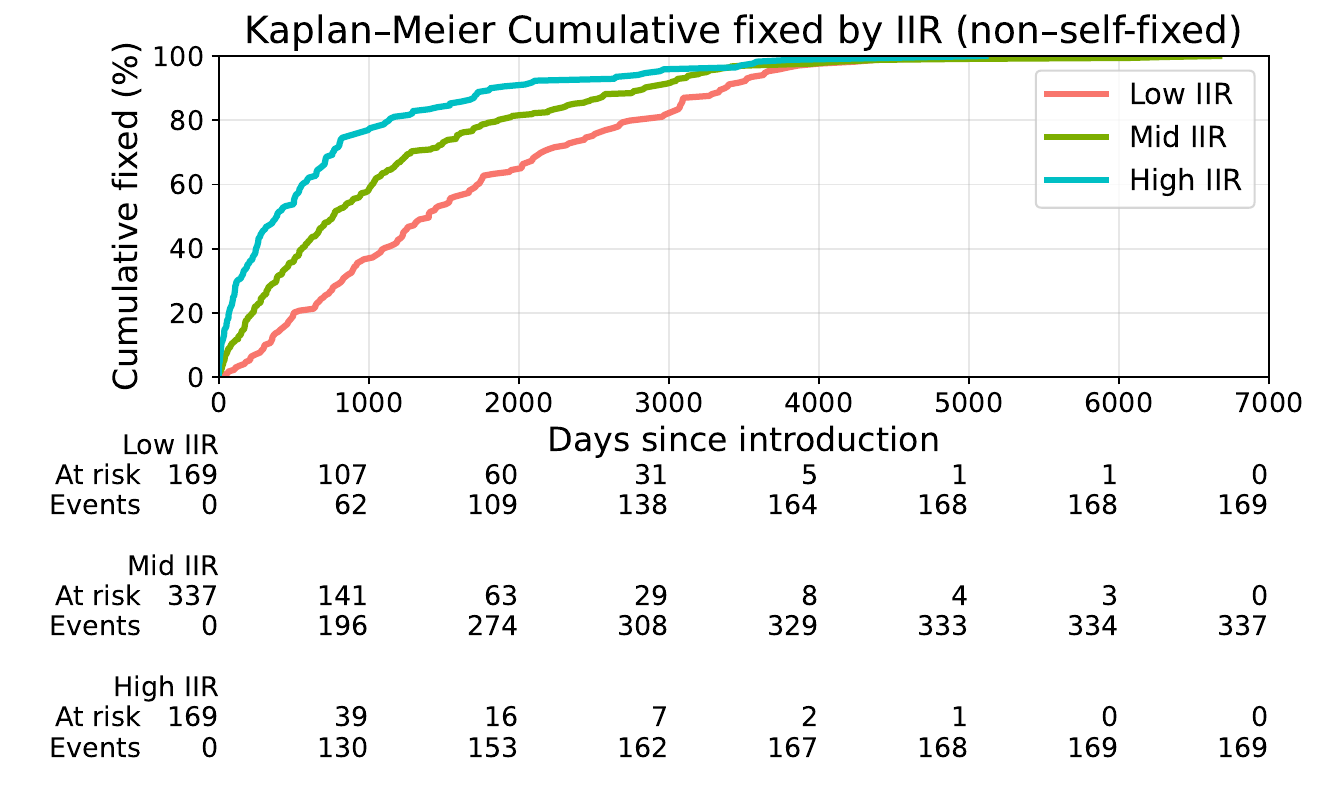}}
    \caption{Kaplan--Meier curves fixed by IIR for non--self-fixed ATD items.}
    \label{fig_KM_IIR_non_self_fixed_full}
 \end{figure*}

We then analyze how these low-, mid-, and high-involvement groups differ in time-to-fix using Kaplan--Meier cumulative-fix curves. \autoref{fig_KM_IIR_non_self_fixed_full} shows the curves stratified by IIR. ATD items with high IIR are repaid the fastest, with the cumulative-fix curve for the high-IIR group consistently above those for the mid- and low-IIR groups. Items with low IIR are repaid substantially more slowly, and the low-IIR curve reaches high cumulative-fix levels later in the timeline.

\begin{figure*}[htpb] 
    \centerline{\includegraphics[trim=0.0cm 0.0cm 0.0cm 0.0cm, clip, width=0.55\textwidth]{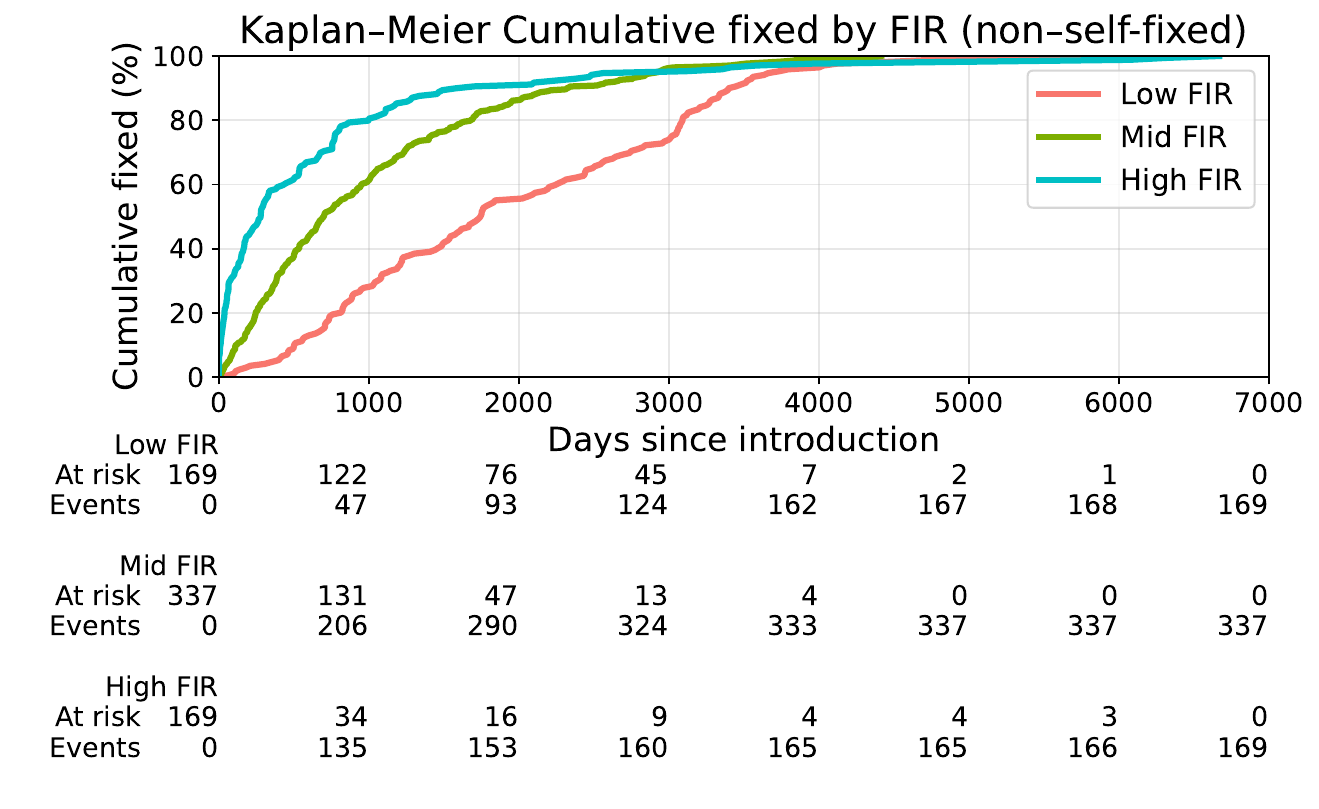}}
    \caption{Kaplan--Meier curves fixed by FIR for non--self-fixed ATD items.}
    \label{fig_KM_FIR_non_self_fixed_full}
 \end{figure*}

\autoref{fig_KM_FIR_non_self_fixed_full} presents the corresponding curves for FIR. Higher fixer involvement is associated with faster repayment, with the low-FIR curve lying below the mid- and high-FIR curves over most of the time range. The high-FIR group reaches a given cumulative-fix level earlier than the low-FIR group, and the mid-FIR group typically follows closely behind the high-FIR group. The mid- and high-FIR curves are relatively close to each other, and both rise more steeply than the low-FIR curve in the initial part of the timeline. Over time, the mid- and high-FIR groups approach similar cumulative-fix levels, while the low-FIR group remains behind and reaches those levels later.

\begin{figure*}[htpb] 
    \centerline{\includegraphics[trim=0.0cm 0.0cm 0.0cm 0.0cm, clip, width=0.55\textwidth]{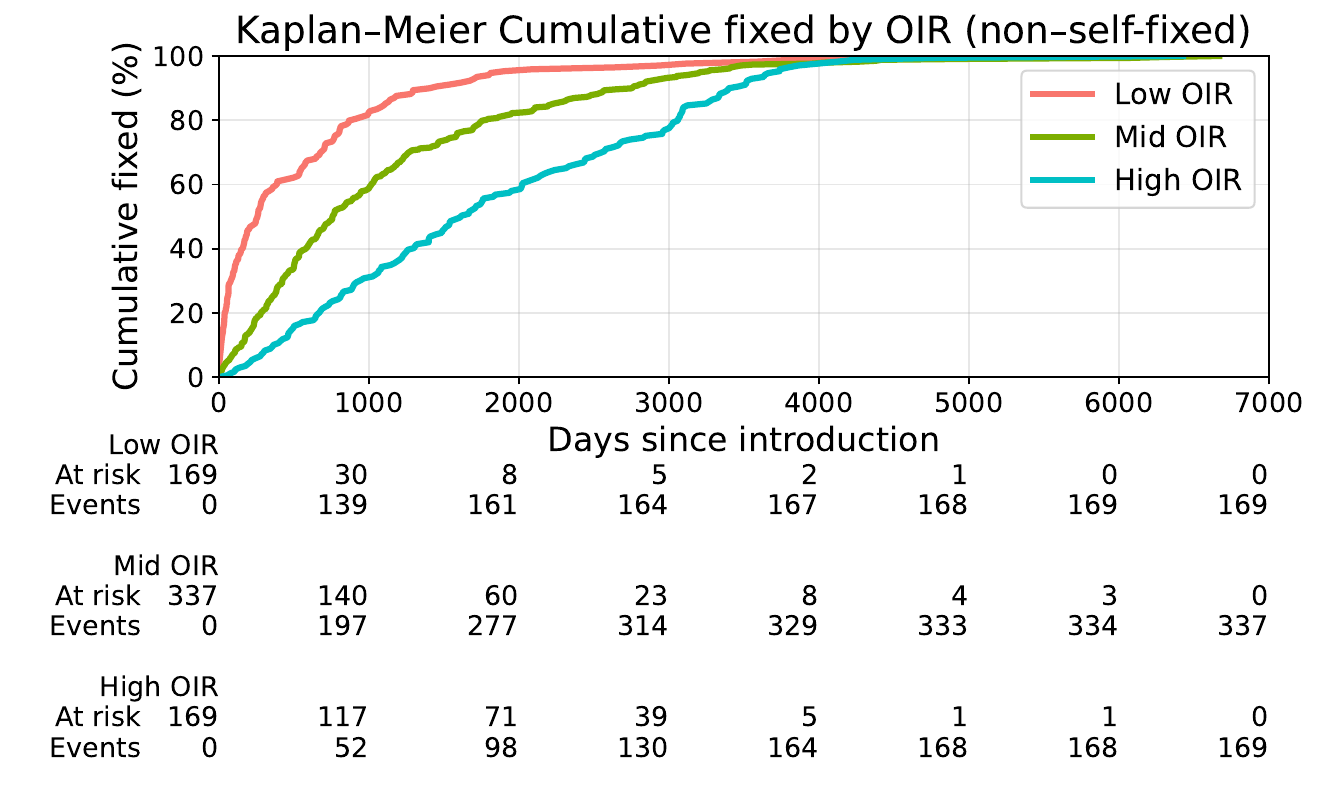}}
    \caption{Kaplan--Meier curves fixed by OIR for non--self-fixed ATD items.}
    \label{fig_KM_OIR_non_self_fixed_full}
 \end{figure*}

Finally, \autoref{fig_KM_OIR_non_self_fixed_full} reports the curves for OIR. In this case, the ordering of the curves is inverted relative to IIR and FIR. ATD items with high OIR, where \enquote{other} developers account for most of the activity, are repaid more slowly, and the high-OIR curve reaches high cumulative-fix levels later than the low- and mid-OIR curves. The low- and mid-OIR curves are very similar to each other and remain above the high-OIR curve throughout most of the observation period. In the early part of the timeline, the low- and mid-OIR groups already show higher cumulative-fix values than the high-OIR group, and this difference persists over time. At later times, the gap between the high-OIR curve and the other two curves remains visible, with a smaller fraction of high-OIR items repaid compared to low- and mid-OIR items at corresponding points in the timeline.

\autoref{tab_rq3_role_desc} shows a clear monotonic pattern in time-to-fix. For IIR, median time-to-fix decreases from approximately 1,399 days in the low-involvement group to 390 days in the high-involvement group; for FIR, it decreases from approximately 1,747 to 279 days. OIR exhibits the opposite pattern, increasing from approximately 259 days in the low group to 1,618 days in the high group.

\begin{table}[t]
  \centering
  \caption{Distribution of role involvement ratios and time-to-fix for non--self-fixed ATD items.}
  \label{tab_rq3_role_desc}
  \tiny
  \begin{tabular}{llrcccccc}
    \toprule
    & & & \multicolumn{3}{c}{Ratio} & \multicolumn{3}{c}{TTF (days)} \\
    \cmidrule(lr){4-6}\cmidrule(lr){7-9}
    Metric & Group & $n$ &
    $\min$ & $\text{median}$ & $\max$ &
    mean & median & stdev \\
    \midrule
    IIR & Low  & 169 & 0.00 & 0.04 & 0.08 & 1,657.94 & 1,399.39 & 1,188.39 \\
        & Mid  & 337 & 0.08 & 0.18 & 0.30 & 1,130.17 &  762.23 & 1,132.42 \\
        & High & 169 & 0.30 & 0.42 & 0.83 &  735.30 &  389.77 &  949.89 \\
    \midrule
    FIR & Low  & 169 & 0.00 & 0.03 & 0.05 & 1,949.41 & 1,746.77 & 1,195.62 \\
        & Mid  & 337 & 0.06 & 0.12 & 0.22 & 1,007.62 &  699.84 &  924.20 \\
        & High & 169 & 0.22 & 0.31 & 0.85 &  688.21 &  278.96 & 1,129.97 \\
    \midrule
    OIR & Low  & 169 & 0.00 & 0.29 & 0.46 &  571.84 &  259.31 &  823.59 \\
        & Mid  & 337 & 0.46 & 0.66 & 0.83 & 1,126.06 &  759.93 & 1,094.90 \\
        & High & 169 & 0.83 & 0.90 & 0.99 & 1,829.61 & 1,618.03 & 1,197.79 \\
    \bottomrule
  \end{tabular}
\end{table}

\autoref{tab_rq3_role_tests} confirms that these differences are statistically significant. For all three involvement ratios, the global log-rank and Kruskal--Wallis tests are significant, and all Bonferroni-adjusted pairwise log-rank comparisons remain significant at $\alpha=0.05$. We therefore \textbf{reject} $H_0$ for all reported comparisons. Overall, greater introducer and fixer involvement is associated with shorter time-to-fix, whereas greater involvement of other developers is associated with longer time-to-fix.

\begin{table}[t]
  \centering
  \small
  \caption{Global and pairwise tests for time-to-fix across role involvement groups of non--self-fixed ATD.
  For pairwise log-rank tests, $p$ is the Bonferroni-adjusted $p$--value; for global log-rank and Kruskal--Wallis tests, $p$ is the raw $p$--value.}
  \label{tab_rq3_role_tests}
  \tiny
  \resizebox{\linewidth}{!}{%
  \begin{tabular}{llccc}
    \toprule
    Metric & Comparison                  & $\chi^{2}$ / $H$ & $p$                 & Sig. ($\alpha=0.05$) \\
    \midrule
    IIR & Global (Low / Mid / High, log-rank) & 52.36  & $4.26\times 10^{-12}$ & Yes \\
        & Low vs Mid (log-rank)               & 18.46  & $5.20\times 10^{-5}$  & Yes \\
        & Low vs High (log-rank)              & 51.57  & $2.08\times 10^{-12}$ & Yes \\
        & Mid vs High (log-rank)              & 16.67  & $1.34\times 10^{-4}$  & Yes \\
        & All groups (Kruskal--Wallis)        & 79.58  & $5.25\times 10^{-18}$ & Yes \\
    \midrule
    FIR & Global (Low / Mid / High, log-rank) & 93.01  & $6.35\times 10^{-21}$ & Yes \\
        & Low vs Mid (log-rank)               & 66.23  & $1.20\times 10^{-15}$ & Yes \\
        & Low vs High (log-rank)              & 77.79  & $3.43\times 10^{-18}$ & Yes \\
        & Mid vs High (log-rank)              & 16.80  & $1.25\times 10^{-4}$  & Yes \\
        & All groups (Kruskal--Wallis)        & 153.80 & $4.00\times 10^{-34}$ & Yes \\
    \midrule
    OIR & Global (Low / Mid / High, log-rank) & 111.52 & $6.09\times 10^{-25}$ & Yes \\
        & Low vs Mid (log-rank)               & 43.85  & $1.07\times 10^{-10}$ & Yes \\
        & Low vs High (log-rank)              & 100.36 & $3.81\times 10^{-23}$ & Yes \\
        & Mid vs High (log-rank)              & 31.89  & $4.89\times 10^{-8}$  & Yes \\
        & All groups (Kruskal--Wallis)        & 139.80 & $4.39\times 10^{-31}$ & Yes \\
    \bottomrule
  \end{tabular}
  }
\end{table}


\begin{tcolorbox}[colback=gray!5!white, colframe=black, title=Summary of Findings (RQ2.1), boxrule=0.01pt, fonttitle=\small, fontupper=\small]
Within non--self-fixed ATD, repayment speed varies systematically with involvement: higher introducer and fixer involvement (IIR/FIR) corresponds to shorter time-to-fix, while higher involvement by other developers (OIR) corresponds to longer time-to-fix, with significant differences across groups and pairwise comparisons.
\end{tcolorbox}

\subsection{RQ3—Differences in development factors and their association with ATD time-to-fix}\label{subsecrq3}

To address RQ3, we first examine how repository-observable development factors differ between self-fixed and non--self-fixed ATD and how they are associated with time-to-fix. We then examine developer project seniority as a complementary factor, focusing on whether developers with longer project tenure are associated with different repayment patterns and shorter time-to-fix. As motivated in Section~\ref{subsubsectionRQ3}, the development-process analysis considers code-change characteristics and development activity.

\begin{figure*}[htpb] 
  \centering
  \subfloat[Change breadth]{%
    \includegraphics[width=0.23\textwidth]{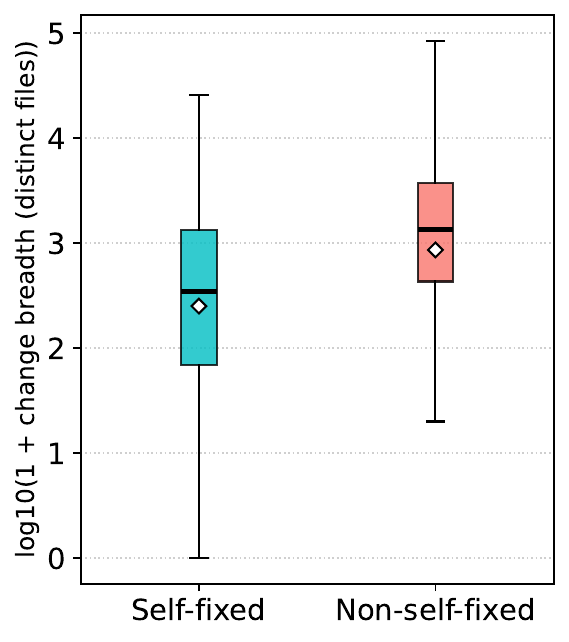}
  }
  \hfill
  \subfloat[Total lines changed]{%
    \includegraphics[width=0.24\textwidth]{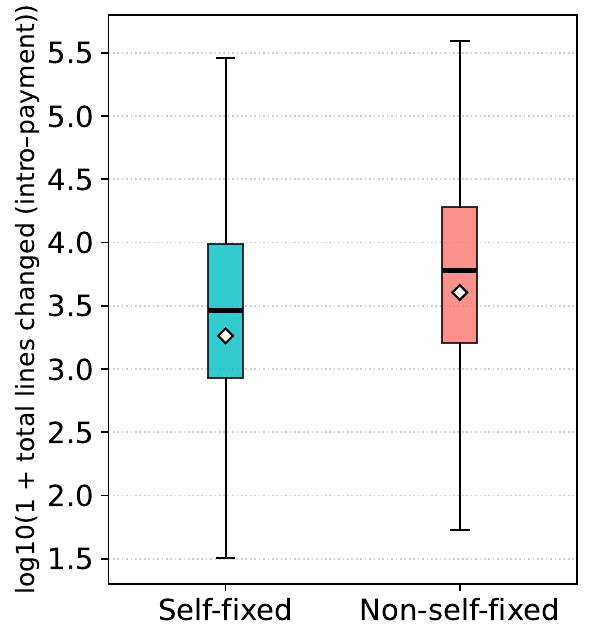}
  }
  \hfill
  \subfloat[Total commits]{%
    \includegraphics[width=0.24\textwidth]{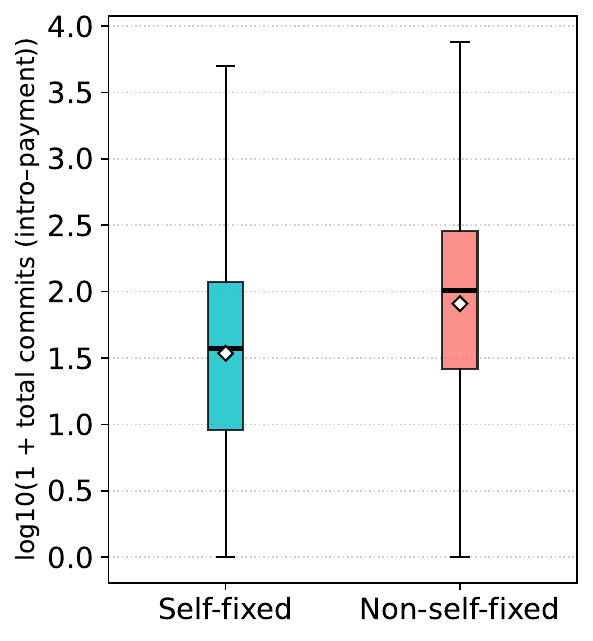}
  }
  \hfill
  \subfloat[Total developers]{%
    \includegraphics[width=0.23\textwidth]{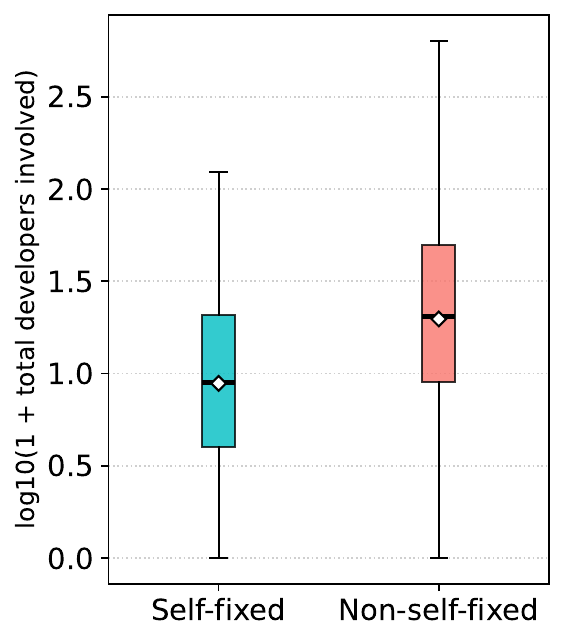}
  }
  \caption{Distribution of code change and development activity metrics for self-fixed vs.\ non--self-fixed ATD items. Values are shown on a $\log_{10}(1+x)$ scale for visual clarity.}
  \label{fig:rq3_boxplots}
\end{figure*}

\autoref{tab:rq3_boxplot_stats} and \autoref{fig:rq3_boxplots} compare the code change characteristics and development activity of self-fixed and non--self-fixed ATD items. We characterize code changes using both \emph{change breadth}, i.e., the number of distinct files modified by commits associated with the tracked ATD files during the introduction--repayment interval, and the total number of changed lines. Development activity is characterized by the number of commits and developers involved during the same interval. For each metric, we use a two-sided Mann--Whitney U test to test the null hypothesis $H_0$ that self-fixed and non--self-fixed ATD items follow the same distribution, with $\alpha=0.05$, and report Cliff's $\delta$ as the effect size. All descriptive statistics and hypothesis tests in \autoref{tab:rq3_boxplot_stats} are computed on the original, non-transformed scale. For visual clarity, the boxplots in \autoref{fig:rq3_boxplots} display the metrics on a $\log_{10}(1+x)$ scale to reduce the influence of extreme values.

At the level of code changes, a pronounced difference emerges for change breadth. Self-fixed ATD has a substantially narrower change breadth than non--self-fixed ATD, with a median of 346 distinct files compared with 1{,}362 files, respectively. The difference is statistically significant ($p=3.73\times10^{-13}$) and has a medium effect size (Cliff's $\delta=-0.36$). We therefore \textbf{reject} $H_0$ for change breadth. As shown in \autoref{fig:rq3_boxplots}\,(a), the distribution of non--self-fixed ATD is shifted towards higher values, indicating that development activity surrounding non--self-fixed ATD tends to span a broader portion of the codebase.

Self-fixed ATD also involves fewer modified lines. The median total number of changed lines between introduction and repayment is 2{,}921 for self-fixed items compared with 6{,}039 for non--self-fixed items. This difference is statistically significant ($p=2.28\times10^{-5}$), although the corresponding effect size is smaller (Cliff's $\delta=-0.21$). We therefore also \textbf{reject} $H_0$ for total lines changed. The difference is visible in \autoref{fig:rq3_boxplots}\,(b), where non--self-fixed ATD is generally associated with larger amounts of code modification.

\begin{table}[htpb] 
  \centering
  \caption{Summary of code change and development activity metrics for self-fixed vs.\ non--self-fixed ATD items. Descriptive statistics are reported on the original scale.}
  \label{tab:rq3_boxplot_stats}
  \tiny
  \resizebox{\linewidth}{!}{%
  \begin{tabular}{llllll}
    \toprule
    Metric & Group & Median & $q_{25}$ & $q_{75}$ & Cliff's $\delta$ / $p$ \\
    \midrule

    \multirow{2}{*}{Change breadth}
      & Self-fixed       &   346.0 &    68.8 & 1{,}329.0 & \multirow{2}{*}{$\delta=-0.36$, $p<0.001$} \\
      & Non--self-fixed  & 1{,}362.0 & 429.8 & 3{,}714.8 & \\

    \midrule

    \multirow{2}{*}{Total lines changed}
      & Self-fixed       & 2{,}921 &   856.3 &  9{,}690.5 & \multirow{2}{*}{$\delta=-0.21$, $p<0.001$} \\
      & Non--self-fixed  & 6{,}039 & 1{,}595.0 & 19{,}078.0 & \\

    \midrule

    \multirow{2}{*}{Total commits}
      & Self-fixed       &  36.0 &  8.0 & 116.3 & \multirow{2}{*}{$\delta=-0.28$, $p<0.001$} \\
      & Non--self-fixed  & 101.0 & 25.0 & 284.0 & \\

    \midrule

    \multirow{2}{*}{Total devs.\ involved}
      & Self-fixed       &  8.0 & 3.0 & 19.8 & \multirow{2}{*}{$\delta=-0.37$, $p<0.001$} \\
      & Non--self-fixed  & 19.5 & 8.0 & 49.0 & \\

    \bottomrule
  \end{tabular}
  }
\end{table}

Beyond the characteristics of the code changes themselves, we also observe marked differences in development activity. Self-fixed ATD involves substantially fewer commits between introduction and repayment than non--self-fixed ATD (median total commits $=36.0$ vs.\ $101.0$; Cliff's $\delta=-0.28$, $p=1.74\times10^{-8}$). We therefore \textbf{reject} $H_0$ for the number of commits. Similarly, self-fixed ATD involves fewer developers overall (median total developers $=8.0$ vs.\ $19.5$; Cliff's $\delta=-0.37$, $p=1.17\times10^{-13}$), corresponding to a medium effect size and indicating more concentrated developer involvement. The distributions in \autoref{fig:rq3_boxplots}\,(c) and \autoref{fig:rq3_boxplots}\,(d) reflect these differences, with non--self-fixed ATD generally involving more commits and a larger number of developers.

Overall, we \textbf{reject} $H_0$ for all four metrics reported in \autoref{tab:rq3_boxplot_stats}. The largest differences are observed for change breadth and the number of developers involved (Cliff's $\delta=-0.36$ and $-0.37$, respectively), whereas the differences in total changed lines and commits are smaller. These findings show that self-fixed ATD is typically associated not only with smaller code changes, but also with development activity that spans fewer files, fewer commits, and fewer developers than non--self-fixed ATD.

We next examine how these code-change and development-activity metrics relate to the observed time-to-fix. To this end, we computed Spearman correlations between time-to-fix (in days) and the four measures reported in \autoref{tab:rq3_boxplot_stats}. Change breadth shows the strongest positive association with time-to-fix ($\rho=0.61$, $p=8.66\times10^{-91}$), indicating that ATD items whose associated development activity spans a broader set of files tend to exhibit longer repayment times. Total lines changed shows a weaker positive association ($\rho=0.36$, $p=4.74\times10^{-29}$). The process-related metrics also exhibit moderate positive associations: the total number of commits ($\rho=0.52$, $p=5.78\times10^{-64}$) and the total number of developers involved ($\rho=0.53$, $p=9.10\times10^{-67}$). Overall, these results indicate that broader and more extensive development activity during an ATD item's lifetime is associated with longer time-to-fix, with change breadth exhibiting the strongest relationship among the examined factors.


\autoref{tab:sigma_desc} summarizes the aggregated change and activity metrics for non--self-fixed ATD items across the three developer-role groups (introducer, fixer, and others).

\begin{table}[htpb]
\centering
\caption{Descriptive statistics (median [$q_{25}$, $q_{75}$]) for aggregated metrics by role in non--self-fixed ATD items ($n=730$).}
\label{tab:sigma_desc}
\resizebox{\linewidth}{!}{%
\begin{tabular}{llll} 
\toprule
Metric & Introducer & Fixer & Others \\
\midrule
$\sum$ lines changed & 1,405.5 [353.25, 4,270.5] & 127.0 [12.0, 620.75] & 2840.0 [448.25, 11,869.75] \\
$\sum$ commits       & 7.0 [2.0, 23.0]         & 2.0 [1.0, 10.0]      & 69.5 [13.0, 240.75] \\
\bottomrule
\end{tabular}
}
\end{table}

To assess whether these differences are systematic within the same ATD items, we apply paired nonparametric tests (\autoref{tab:sigma_tests}).

\begin{table}[htpb]
\centering
\caption{Paired statistical tests for aggregated metrics by role in non--self-fixed ATD items ($n=730$). Global test: Friedman. Post-hoc: Wilcoxon signed-rank with Holm correction; effect size: RBC.}
\label{tab:sigma_tests}
\resizebox{\linewidth}{!}{%
\begin{tabular}{lccccc}
\toprule
Metric & $\chi^2(2)$ & $p$ & I--F ($p_h$, RBC) & I--O ($p_h$, RBC) & F--O ($p_h$, RBC) \\
\midrule
$\sum$ lines changed
& 512.548 & $5.03\times10^{-112}$
& $6.99\times10^{-66}$, 0.755
& $1.73\times10^{-24}$, $-0.447$
& $2.93\times10^{-83}$, $-0.853$ \\
$\sum$ commits
& 712.611 & $1.81\times10^{-155}$
& $1.87\times10^{-12}$, 0.333
& $6.83\times10^{-93}$, $-0.907$
& $1.48\times10^{-100}$, $-0.943$ \\
\bottomrule
\end{tabular}
}
\end{table}

Tables~\ref{tab:sigma_desc} and \ref{tab:sigma_tests} show that Others account for the greatest aggregated activity, followed by the Introducer, whereas the eventual Fixer contributes the least. The Friedman tests and all Holm-adjusted pairwise comparisons are significant for both changed lines and commits. The effect sizes show that introducers contribute more than fixers, whereas Others contribute more than both introducers and fixers. Thus, development activity during non--self-fixed ATD lifetimes is comparatively distributed beyond the two focal developers, with the eventual fixer accounting for the smallest aggregated activity.

\begin{figure*}[htpb] 
  \centering
  \centerline{\includegraphics[trim=0.0cm 0.0cm 0.0cm 0.0cm, clip, width=0.75\textwidth]{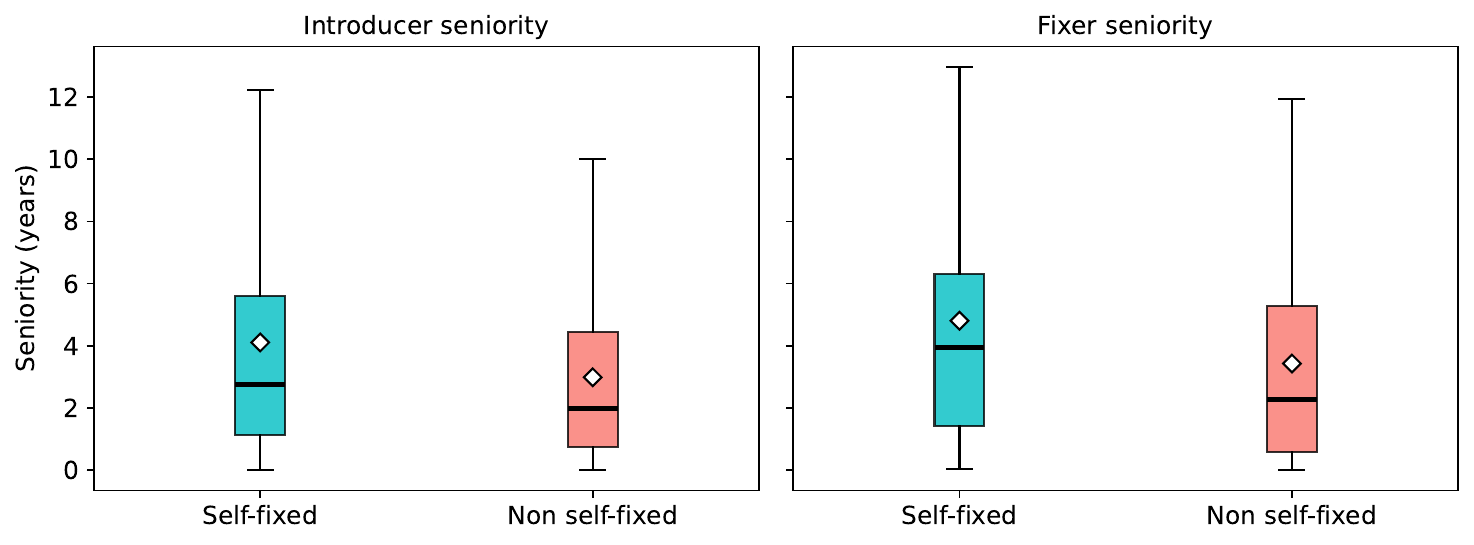}}
  \caption{Developer seniority (in years since first project commit) for self-fixed vs.\ non--self-fixed ATD items, separated by introducer and fixer roles.}
  \label{fig:rq3_seniority}
\end{figure*}

Following Eyolfson et al.\ \cite{eyolfson2014correlations}, we analyze the project seniority of developers who introduce and repay each ATD item. \autoref{fig:rq3_seniority} visualizes the distributions of introducer and fixer seniority for self-fixed and non--self-fixed ATD. Self-fixed items tend to be introduced by slightly more senior developers than non--self-fixed items (median introductory seniority $=2.76$ vs.\ $1.99$ years; Cliff's $\delta=0.158$, $p=1.51\times10^{-3}<0.05$). A similar but more pronounced difference emerges for the developers who repay ATD; self-fixed items are repaid by more senior developers (median repayment seniority $=3.93$ vs.\ $2.26$ years; Cliff's $\delta=0.218$, $p=1.11\times10^{-5}<0.05$). 


For self-fixed ATD, the introducer and fixer are the same developer. The two self-fixed distributions in \autoref{fig:rq3_seniority} therefore represent two different points in the same developer's project tenure rather than two different individuals. Because introducer seniority is measured at introduction and fixer seniority at the later repayment event, the same developer is necessarily more senior at repayment. Consequently, the upward shift of the self-fixed fixer distribution relative to the self-fixed introducer distribution should not be interpreted as evidence that increasing seniority itself leads to faster repayment.

To examine more directly whether prior project experience is associated with repayment speed, we additionally consider the experience already accumulated by the eventual fixer when the ATD was introduced. We define \emph{fixer prior seniority} as the elapsed time between the fixer's first recorded commit in the project and the ATD introduction date. Among eventual fixers who were already active when the ATD was introduced, fixer prior seniority shows no significant monotonic association with time-to-fix (Spearman's $\rho=-0.039$, $p=0.371$). Thus, within the analyzed data, the exact amount of prior project tenure does not reveal a clear seniority level associated with faster ATD repayment.

Because tenure length alone does not show a clear relationship with repayment time, we further distinguish whether the eventual fixer was already active in the project when the ATD was introduced or joined the project only afterward. This distinction captures temporal continuity of project participation. Among the ATD items for which the fixer's project-entry date could be determined, items whose eventual fixer was already active at introduction exhibit a substantially shorter median time-to-fix than items repaid by developers who joined the project after the ATD had been introduced (291.02 vs.\ 1,227.96 days; Mann--Whitney $U=41{,}300$, $p=3.54\times10^{-47}$). This result indicates that prior participation in the project at the time the debt emerged is more strongly associated with repayment time than the exact length of fixer seniority. It should not be interpreted causally, because later-joining fixers may encounter older or otherwise different ATD items.

Beyond the relationship between developer experience and repayment time, we examine whether developer characteristics are associated with the likelihood of self-fixing using the GLMM specified in Subsection~\ref{subsecrq3}. \autoref{tab_glmm_self_fixed} presents the model results.

\begin{table}[htpb] 
  \centering
  \caption{Logistic GLMM for the likelihood that an ATD item is self-fixed.
  Positive coefficients increase the odds of self-fixing.}
  \label{tab_glmm_self_fixed}
  \resizebox{\linewidth}{!}{%
  \begin{tabular}{lrrrrrr}
    \toprule
    Predictor & $\beta$ & SE & OR & 95\% CI (OR) & $p$ & $p$ vs.\ 0.05 \\
    \midrule
    Intercept                          & -1.80 & 0.22 & 0.17 & [0.11, 0.25]    & $< 0.001$            & $< 0.05$ \\
    Introducer seniority (z)           &  0.37 & 0.12 & 1.45 & [1.15, 1.82]    & $1.52\times 10^{-3}$ & $< 0.05$ \\
    Introducer total commits (z)       &  1.48 & 0.54 & 4.39 & [1.51, 12.74]   & $6.60\times 10^{-3}$ & $< 0.05$ \\
    Introducer total files touched (z) & -1.29 & 0.57 & 0.28 & [0.09, 0.85]    & $2.47\times 10^{-2}$ & $< 0.05$ \\
    \bottomrule
  \end{tabular}
  }
\end{table}

The GLMM results in \autoref{tab_glmm_self_fixed} show that all three developer-related predictors are statistically significant at $\alpha=0.05$. Introducer seniority is positively associated with the likelihood of self-fixing (OR $=1.45$), indicating that a one-standard-deviation increase in seniority is associated with approximately 45\% higher odds of self-fixing. Introducer commit activity shows a stronger positive association (OR $=4.39$), whereas the number of distinct files touched by the introducer is negatively associated with self-fixing (OR $=0.28$). The random-intercept variances indicate greater heterogeneity between introducers (variance $=0.80$) than between projects (variance $=0.13$).

The RQ3 results reveal complementary development- and developer-level patterns. Self-fixed ATD is associated with narrower change breadth, smaller change size, fewer commits, and fewer involved developers. Self-fixed items are also more frequently associated with introducers having greater project seniority and commit activity. However, the exact amount of fixer seniority accumulated before ATD introduction is not significantly associated with time-to-fix. A clearer distinction emerges from temporal project participation: ATD is repaid substantially sooner when the eventual fixer was already active in the project at introduction than when the fixer joined afterward.

\begin{tcolorbox}[colback=gray!5!white, colframe=black, title=Summary of Findings (RQ3), boxrule=0.01pt, fonttitle=\small, fontupper=\small]

Self-fixed ATD is associated with narrower change breadth, smaller change size, fewer commits, and fewer involved developers than non--self-fixed ATD. Among these development factors, change breadth shows the strongest positive association with time-to-fix. A further distinction in repayment time emerges from project involvement at ATD introduction: ATD items are repaid substantially sooner when the eventual fixer was already active in the project when the debt was introduced. The GLMM further shows that introducer seniority and contribution characteristics are associated with the likelihood of self-fixing.

\end{tcolorbox}

\section{Discussion}\label{secDiscussion}
In this section, we interpret the empirical findings in Section~\ref{secResults} and discuss implications for researchers and practitioners.

\subsection{Implications for researchers}
\paragraph{RQ1—Prevalence of self-fixing in ATD repayment\\}
Our results show that only a minority of ATD items are repaid by their introducers, whereas most are repaid by other developers. This suggests that ATD repayment often occurs when the repayment is handled by developers other than the one who introduced it. This pattern is consistent with Tan et al.~\cite{tan2020empirical}, who also reported that many technical debt items are repaid by developers other than the introducers. However, our self-fixing rate for ATD is lower (18.5\%) than the self-fixing rate they reported for other types of technical debt in Java projects (33.26\%), and higher rates are reported for Python projects (65.54\%)~\cite{tan2022does}. These differences suggest that self-fixing prevalence can vary by debt type and project context, and it should not be treated as a universal baseline. Since ATD is considered the most critical form of technical debt \cite{besker2017impact,verdecchia2018architectural}, these differences matter when interpreting ATD repayment dynamics and time-to-fix.

An additional factor to consider is the difference in the roles and knowledge required for repayment. Prior studies examine several TD types whose repayment may involve a broad range of developers, whereas ATD often concerns system-level design decisions and cross-component dependencies that require architectural or component-specific knowledge. Consequently, the developer who introduces an ATD item is not necessarily the developer who later has the knowledge or responsibility needed to repay it. This distinction may contribute to the lower self-fixing rate observed for ATD. However, because our repository data do not identify formal roles such as software architect, maintainer, or developers responsible for specific components, this relationship is not directly tested in our study.

The low self-fixing rate may also be related to the temporal nature of ATD repayment. Architectural knowledge can erode over time due to knowledge decay~\cite{borrego2019towards} and developer turnover~\cite{foucault2015impact}, which can make repayment more dependent on other developers. Therefore, researchers should consider ecosystem and governance factors when analyzing ATD repayment and time-to-fix. Future work should replicate ATD-focused analyses in other language ecosystems, including \texttt{Python} projects, to test whether these patterns remain stable across contexts.

\paragraph{RQ1.1—Distribution of contribution shares on ATD-affected files\\}
Our results show that contribution activity in self-fixed ATD is more concentrated around the self-fixer under both commit- and change-size-based measures, with a stronger concentration in changed lines. In contrast, non--self-fixed ATD exhibits a more distributed pattern across the introducer, eventual fixer, and other developers. This indicates that commit frequency and change magnitude capture complementary aspects of developer involvement and that commit counts alone may not fully characterize contribution concentration. 

Although self-fixed ATD also exhibits shorter time-to-fix in RQ2, the present analysis does not establish whether a higher degree of self-fixer contribution concentration is itself associated with earlier repayment. These findings should therefore not be interpreted as evidence that greater contribution concentration leads to faster repayment. Future work could directly examine this relationship by relating item-level contribution-concentration measures to time-to-fix across additional projects.

\paragraph{RQ2—Time-to-fix comparison between self-fixed and non--self-fixed ATD\\}
While we observe that self-fixed ATD is repaid faster than non--self-fixed ATD, this difference in time-to-fix does not yet reveal the underlying technical mechanisms. Future work should therefore connect repayment dynamics to structural and effort-related properties of the affected code by examining changes in complexity and coupling measures across the introduction and repayment phases. In particular, researchers can analyze cyclomatic (or cognitive) complexity as a proxy for local change effort and risk and FAN-IN and FAN-OUT as proxies for coupling and modularity. Such analyses can show whether faster repayment reflects smaller, localized fixes or more effective architectural improvements and can reduce confounding by controlling for baseline complexity and coupling at introduction. Extending survival models to include these covariates and conducting stratified analyses by ATD type would provide a more mechanistic explanation for why self-fixing is associated with shorter time-to-fix.

\paragraph{RQ2.1—Non--Self-fixed ATD and Developers' Involvement\\}
Our results show that non--self-fixed ATD items can follow different involvement patterns, as captured by the role-specific Involvement Ratios (IIR, FIR, and OIR). This implies that researchers should not treat non--self-fixed ATD as a single category when studying time-to-fix. Instead, future studies should incorporate IIR, FIR, and OIR directly in their models of time-to-fix, rather than relying only on coarse measures such as the number of developers or commits. Researchers can also extend this analysis by computing IIR, FIR, and OIR using different contribution measures. These can be derived from existing ATD lifecycle studies, for example, the total lines changed or code churn, to test whether the same involvement patterns hold when contribution magnitude is quantified by change size, rather than by commit counts. Finally, replications across additional projects and ecosystems are needed to assess whether the relationships between IIR/FIR/OIR and time-to-fix remain stable across contexts and to identify the project characteristics that amplify or weaken these effects.

\paragraph{RQ3—Differences in development factors and their association with ATD time-to-fix\\} Our results show that both code-change characteristics and development activity are associated with ATD repayment time. Among the examined factors, change breadth exhibits the strongest positive association with time-to-fix ($\rho=0.61$), followed by the number of involved developers ($\rho=0.53$), the number of commits ($\rho=0.52$), and change size ($\rho=0.36$). This suggests that the extent to which ATD-associated development activity spreads across the codebase may be particularly informative when characterizing prolonged repayment. However, these relationships are observational and should not be interpreted as evidence that broader changes cause longer repayment.

In addition, the association between seniority, contribution activity, and the likelihood of self-fixing is consistent with the possibility that project-specific knowledge plays a role in ATD repayment. One possible explanation is that developers with greater project-specific experience or activity may possess more contextual knowledge of the affected subsystems. However, our repository data do not directly measure architectural knowledge, formal responsibility, or decision-making authority, so this mechanism cannot be established from the present analysis. Moreover, ATD items are repaid substantially sooner when the eventual fixer was already active in the project at the time of debt introduction, suggesting that prior project involvement may be more informative for repayment speed than tenure length alone.


Future studies should explicitly model change breadth, change size (e.g., code churn), the number of commits, and the number of involved developers during the introduction--repayment interval. Multivariate and survival models incorporating these factors could help determine whether each provides explanatory information beyond the others and assess whether the observed relationships generalize across projects and ATD types.

\subsection{Implications for practitioners}
\paragraph{RQ1—Prevalence of self-fixing in ATD repayment\\}
Given that most ATD items are repaid by developers other than their introducers, projects should preserve the architectural context needed for future repayment rather than relying on the continued availability of the original developer. ATD issues should document the architectural decision or trade-off, the affected components and constraints, the conditions under which repayment is expected, and the intended acceptance criteria. Projects can also assign a maintainer or responsible developer to architecture-critical components to periodically review open ATD items and ensure that repayment is explicitly planned.

\paragraph{RQ1.1—Distribution of contribution shares on ATD-affected files\\}
When development activity on ATD-affected files is distributed across several contributors, teams should make repayment responsibility explicit. In particular, the number of involved developers and the distribution of activity among the introducer, fixer, and other contributors can be used to identify items that require greater coordination. For such items, teams can appoint a responsible coordinator, define the scope of the architectural change, and use consistent reviewers for architecture-critical modifications.

\paragraph{RQ2—Time-to-fix comparison between self-fixed and non--self-fixed ATD\\}
Our results show that non--self-fixed ATD is not only common but also takes much longer to repay than self-fixed ATD. This time-to-fix gap means that non--self-fixed ATD can remain unresolved for longer, extending the period during which the architectural compromise remains present in the system. Therefore, teams should treat ATD as a higher-priority maintenance item than technical debt types other than ATD and manage it with more explicit planning and follow-up, especially when repayment is likely to be carried out by developers other than the introducer. This is especially important because ATD is perceived as the most negatively impactful type of technical debt in daily work \cite{besker2017impact} and often affects multiple subsystems and requires more coordination \cite{samarthyam2016refactoring}.

In practice, the time-to-fix gap suggests two actions. First, when the introducer is still active, teams should review ATD issues early and plan repayment soon, while the design context is still clear. Second, if an ATD item will be repaid by someone else (or if the introducer becomes inactive), the team should actively plan the work rather than leave it in the backlog. This means assigning a responsible developer, defining the scope and acceptance criteria, and recording the architectural rationale needed to complete the repayment.

\paragraph{RQ2.1—Non--Self-fixed ATD and Developers' Involvement\\}
When feasible, the introducer can remain involved in design discussions, repayment planning, or code review even when another developer performs the final repayment. When such involvement is not possible, architectural knowledge should be transferred through lightweight documentation, such as structured issue templates or architecture decision records. At minimum, this documentation should capture why the debt was introduced, the architectural constraints involved, the affected components, and the intended repayment strategy.


Finally, ATD items where \enquote{Others} dominate activity deserve attention because they appear more prone to extended repayment delays. Teams can respond by explicitly assigning accountability, prioritizing such items in architectural triage, and scheduling refactoring windows or milestone-based cleanup to prevent prolonged deferral and the resulting architectural erosion \cite{li2022understanding}. Involving experienced contributors early may provide additional architectural context during repayment, although the effectiveness of such an intervention requires further empirical evaluation.

\paragraph{RQ3—Development factors and ATD time-to-fix\\}
Our findings show that ATD items whose associated development activity spans a broader set of files tend to exhibit longer repayment times. Repayment time is also positively associated with the number of involved developers, the number of commits, and the amount of code changed. Teams can therefore monitor not only how many developers and commits accumulate around an ATD item, but also how broadly the associated changes spread across the codebase. A rapidly expanding change breadth may indicate that repayment is becoming increasingly distributed and may warrant clearer ownership, scope control, and architectural coordination.

These repository-derived indicators could also complement TD management tools with a role-aware view of ATD repayment. In addition to identifying and prioritizing debt, such tools could report the introducer and fixer, the number of involved developers and commits, and the distribution of activity across developer roles. This information could support architectural triage by highlighting ATD items involving more distributed development activity and therefore requiring explicit coordination.

From a team-organization perspective, our results support clear repayment accountability and focused contributor groups, but they do not provide evidence for prescribing a fixed fixer-to-introducer ratio. The observed self-fixed and non--self-fixed proportions describe how ATD items were repaid, rather than the workload capacity of individual developers. An appropriate staffing ratio would additionally depend on the number and complexity of ATD items, repayment effort, developer availability, and component-specific expertise. Thus, our findings support assigning clear ownership of repayment rather than prescribing a specific number of fixers per introducer.


\subsection{Threat to validity}
This section discusses potential threats to the interpretation and generalizability of our findings.

\subsubsection{Construct validity}
We infer introduction commits using an SZZ-style strategy that traces backward from the repayment commit. Candidate commits are recovered via fine-grained diffs and blame, then ranked by the number of blamed lines attributable to the remediation; ties are resolved by selecting the earliest candidate. This inference can be threatened by refactorings, file moves/renames, formatting-only edits, or repeated edits to the same lines, which may distort blame and shift the inferred introduction point. To mitigate these threats, we use a systematic tracing procedure implemented with PyDriller for fine-grained diffs and blame-based candidate recovery, which is widely used in software maintenance research~\cite{le2021deepcva,iannone2023rubbing,wehaibi2016examining,wen2022quick}, and manually verify a randomly selected 5\% sample of items (see Section~\ref{subsecRM}), achieving 93.3\% accuracy for the inferred introduction commits in the inspected sample.

We also note a construct-level limitation when comparing our results with prior work (e.g., Tan et al.~\cite{tan2022does}). Their study relies on SonarQube-based TD detection, whereas we use manual labeling of ATD items. This difference in operationalization can affect prevalence and repayment estimates across studies. SonarQube-based detection can produce false positives~\cite{lenarduzzi2020sonarqube}. To avoid this source of measurement error, our study relies on manual labeling rather than SonarQube. We therefore interpret cross-study comparisons cautiously and rely primarily on within-dataset contrasts where detection and measurement procedures are consistent.

In addition, the main analysis combines True-ATD and Weak-ATD items. While True-ATD represents cases with unambiguous evidence of architectural technical debt, Weak-ATD represents more borderline cases with lower classification confidence. We acknowledge that combining these categories may introduce uncertainty into the results. To assess the impact of this decision, we repeated the main analyses using only True-ATD items. The resulting patterns were consistent with those of the main analysis, indicating that the principal findings are robust to the exclusion of Weak-ATD items. The complete True-ATD-only robustness analysis is provided in the replication package.

\subsubsection{External validity}
Our dataset is drawn from 10 Apache open-source projects and includes only ATD items that can be traced from Jira to remediation commits via commit-message references; therefore, the results may not generalize to other ecosystems (e.g., GitHub Issues-only workflows) or industrial settings. While our dataset is relatively small, consisting of a sample of about 1{,}100 ATD items, of which 896 can be reliably tracked to remediation commits, this scale can be considered modest compared to large-scale mining datasets and is consistent with prior work in many software engineering studies~\cite{wen2019exploring,al2024detecting,prenner2021making,tsantalis2018accurate}.

\subsubsection{Internal validity}
In some projects, developers may remain responsible for components they previously modified, which can increase the likelihood of self-fixing independently of their retained architectural knowledge. Conversely, organizational structures that separate development, architecture, testing, and maintenance responsibilities may make non--self-fixed repayment more likely by design. Our repository data do not provide complete information about formal task assignments or team roles; therefore, we cannot distinguish whether self-fixing results from retained contextual knowledge, component ownership, work allocation, or a combination of these factors. We consequently interpret the observed differences as associations between repayment mode and repayment dynamics rather than causal effects of self-fixing. Alternative explanations are possible; for example, more difficult ATD items may both take longer to repay and involve more developers and commits. To mitigate threats to internal validity, we use comparative, distribution-aware statistical designs well-suited to skewed time-to-fix data, including non-parametric tests and survival analysis. We therefore interpret the results as repayment dynamics and correlates, not as evidence of causal impact.

\section{Conclusion}\label{secConclusion}
This paper investigated the relationship between \enquote{who} repays ATD and \enquote{how long} repayment takes, using a large-scale empirical dataset of repaid and traceable ATD items from ten Apache projects. We distinguished \textit{self-fixed} ATD from \textit{non--self-fixed} ATD and examined repayment prevalence, time-to-fix differences, and development factors associated with repayment latency.

Our results show that self-fixing is uncommon for ATD: most ATD items are repaid by developers other than their introducers. Moreover, non--self-fixed ATD exhibits a substantially longer time-to-fix than self-fixed ATD. This pattern is consistent with the possibility that repayment by another developer may involve additional knowledge transfer or coordination, although these mechanisms are not directly measured in our repository data. Finally, we find that repayment latency is positively associated with several repository-observable development factors. Change breadth exhibits the strongest association with time-to-fix, followed by the number of involved developers and commits, while change size shows a weaker but still significant association. These findings indicate that prolonged ATD repayment is associated not only with the magnitude of code modification, but also with how broadly development activity spreads across files and developers.

Future work should replicate these findings in other ecosystems and industrial settings and further examine whether the same relationships hold by combining repository mining with qualitative evidence, such as design discussions and architecture decision records.

\section*{Acknowledgment}
This work was financially supported by the Indonesian Education Scholarship (BPI) from the Center for Higher Education Funding and Assessment (PPAPT), Indonesia Endowment Fund for Education (LPDP), and the Ministry of Higher Education, Science, and Technology of the Republic of Indonesia.

\section*{Author Contributions}
\textbf{Edi Sutoyo}: Conceptualization; Data curation; Methodology; Writing - original draft preparation; Visualization. \textbf{Paris Avgeriou}: Conceptualization; Supervision; Writing - review and editing. \textbf{Andrea Capiluppi}: Conceptualization; Supervision; Writing - review and editing.

\section*{Data Availability}
The data associated with this study are publicly available online in the replication package.\footnote{\url{https://github.com/edisutoyo/self-fixed-atd}}

\section*{Declarations}

\paragraph{\textbf{Conflicts of Interests}\\}
The authors have no competing interests to declare that are relevant to the content of this article.

\paragraph{\textbf{Ethical Approval}\\}
Not applicable.

\paragraph{\textbf{Clinical Trial Number}\\}
Not applicable.

\paragraph{\textbf{Informed Consent}\\}
Not applicable.

\paragraph{\textbf{Funding}\\}
This work was financially supported by the Indonesian Education Scholarship (BPI) from the Center for Higher Education Funding and Assessment (PPAPT), Indonesia Endowment Fund for Education (LPDP), and the Ministry of Higher Education, Science, and Technology of the Republic of Indonesia.

\tiny
\bibliographystyle{spmpsci}

\bibliography{main-bibliography}

\end{document}